\documentclass[11pt]{article}
\usepackage{amsmath}
\usepackage{amsfonts}
\usepackage{amssymb}
\usepackage[dvips]{epsfig}
\usepackage{graphicx}
\usepackage{mathrsfs}
\usepackage{color}
\usepackage{feynmf}

\begin{document}

\title{Path Integral Quantization of Generalized Quantum Electrodynamics}
\author{R. Bufalo$^{1}$\thanks{%
rbufalo@ift.unesp.br}~, B.M. Pimentel$^{1}$\thanks{%
pimentel@ift.unesp.br}~ and G. E. R. Zambrano$^{2}$\thanks{%
gramos@udenar.edu.co}~ \\
\textit{{$^{1}${\small Instituto de F\'{\i}sica Te\'orica (IFT/UNESP), UNESP
- S\~ao Paulo State University}}} \\
\textit{\small Rua Dr. Bento Teobaldo Ferraz 271, Bloco II Barra Funda, CEP
01140-070 S\~ao Paulo, SP, Brazil}\\
\textit{{$^{2}${\small Departamento de F\'{\i}sica, Universidad de Nari\~{n}o%
}}} \\
\textit{{\small Clle 18 Cra 50, San Juan de Pasto, Nari\~{n}o, Colombia}}}
\date{}
\maketitle

\begin{abstract}
In this paper, a complete covariant quantization of generalized
electrodynamics is shown through the path integral approach. To this goal, we first studied the
hamiltonian structure of system following Dirac's methodology and, then, we
followed the Faddeev-Senjanovic procedure to obtain the transition amplitude.
The complete propagators (Schwinger-Dyson-Fradkin equations) of the correct
gauge fixation and the generalized Ward-Fradkin-Takahashi identities are
also obtained. Afterwards, an explicit calculation of one-loop approximation
of all Green's functions and a discussion about the obtained results are
presented.
\end{abstract}



\section{ Introduction}

The results that have been obtained for known theories using available theoretical tools
are very impressive: the agreement of $QED_{4}$ with experiments,
predictions of standard model and $QCD_{4}$, and so many others. A point
that warrants comment is the effectiveness of such theories up to a
determined energy scale. Usually, a physics problem involves widely
separated energy scales; this allows us to study the low-energy dynamics
independently of the details of the high-energy interactions. The main idea
is to identify those parameters that are very large (small) compared to
the relevant energy scale of the physical system and let them go to infinity
(zero). This provides a sensible approximation to the problem, which can
always be improved by taking into account the corrections induced by the
neglected energy scales as small perturbations. Effective theories constitute the
appropriate theoretical tools to describe low-energy physics, where low is
defined with respect to some energy scale. This idea of effective theories
was proposed by Weinberg \cite{1}.

The set of higher-order theories belongs to such effective theories. As it is known, the majority of physical systems described by
Lagrangians depends, at most, on first-order derivatives. However, with the
first development in formal aspects of higher-order derivative Lagrangians
in classical mechanics by Ostrogradski \cite{2}, a new field of research was
opened.

The branch of higher-order derivative theories becomes very interesting, due to
the fact that these additional terms are constructed in such way so as to preserve
the original symmetries of problem. As a remark, it is important to say that
this kind of theory has been shown to be a powerful method for
consistent regularization of the ultraviolet divergences of gauge-invariant
and supersymmetric theories \cite{3}. Also, the use of higher derivative
terms becomes interesting regulatorr, by the fact that it improves the
convergence of the Feynman diagrams \cite{4}.

More examples of systems treated with high-order Lagrangians that we can
mention are: the study of the problem of color confinement on the infrared
sector of $QCD_{4}$ \cite{5}, the attempts to solve the problem of
renormalization of the gravitational field \cite{6}, and a generalization of
Utiyma's theory to second-order theories \cite{7}. Although all these works
improve the use of higher-order terms, the ones that most contributed to show
the effectiveness of such terms in field theory was the contributions of
Bopp \cite{8}, and Podolsky and Schwed \cite{9}, where they proposed a
generalization of the Maxwell electromagnetic field. They wanted to get rid
of the infinities of the theory, such as the electron self-energy ($r^{-1}$
singularity) and the vacuum polarization current present on the Maxwell
theory. The modification suggested by Podolsky and Schwed handle these
unsolved problems and, also, gives a positive definite energy in the
electrostatic case; also, as showed by Frenkel \cite{10}, it gives the correct
expression for the self-force of charged particles. In \cite{7}, it was
shown that the Podolsky Lagrangian is the only possible generalization of
Maxwell electrodynamics that preserves invariance under $U(1)$.

On theoretical and experimental framework, efforts have been made to determine an
upper-bound value for the mass of the photon \cite{11}, the existence
of a massive sector being a prediction of generalized electrodynamics. Along this line
of thought, we believe that a way to set limits over Podolsky
parameter will be to study the Podolsky's photons interacting with standard
model particles, and compare the obtained results with high-energy
experiments. This idea and other purposes led Podolsky and some of his
students to study the interaction of electrons with the Podolsky photons, which
they called generalized quantum electrodynamics ($GQED_{4}$) \cite{12}.
Among the points dealt with in their thesis, the most interesting was the
calculation of electron self-energy at a one-loop approximation. They expected
that the contribution of massive photons lead to a finite result;
nevertheless, in the end, they found, as in the usual $QED_{4}$, a divergent
expression. Analyzing, now, the thesis results, we found a mistake in
their treatment of theory, i.e., the choice of usual Lorenz gauge condition.
However, this analysis was only possible due to the contribution of Galv\~{a}%
o and Pimentel \cite{13}, which gives the first consistent quantization to
Podolsky theory, where Dirac Hamiltonian formalism \cite{14} was used with
the correct choice of gauge condition, which they called the generalized Lorenz
gauge condition. Also, they showed that, different from the usual Lorenz
condition, the generalized one fulfills all the requirements for a good
choice of gauge condition on the context of Podolsky theory. Indeed, one
of the aims of this paper is to quantize $GQED_{4}$, now, in the generalized
Lorenz gauge condition. The Podolsky electrodynamics, by itself, takes
account of several classical problems of Maxwell's theory, and it should be
expected that the addition of Podolsky term into the $QED_{4}$ Lagrangian
with an appropriated gauge choice should give rise to interesting results.

Based on all these facts pointed out above, we can conclude that higher-order
theories deserve a deeper investigation. Therefore, this paper is intended to
give a correct and transparent quantization of $GQED_{4}$, the interaction
of electrons with Podolsky's photons in four dimensional space-time. To
improve our understanding of the features of the $GQED_{4}$, we proceeded to calculate
the radiative corrections of Green's functions. The main results of
the paper will be closed formulas to the complete propagators and vertex
function by functional methods (for a excellent review see \cite{15}), and
it turns out that, with the correct gauge choice, the electron and vertex
self-energy functions are finite at $e^{2}$-order approximation.

The work is organized as follow. In Sec.2, we present a brief study of
canonical structure of the theory and, then, construct the transition
amplitude by the Faddeev-Senjanovic procedure \cite{16}, which we believe is the
most appropriated for our interest. In Sec.3, we introduce the generating
functional, which will generate all the Green's functions, photon and
electron propagator, and vertex function; also, through it, we will derive the
generalized Ward-Fradkin-Takahashi identities in Sec.4. In Sec.5, we
evaluate and discuss the self-energy functions of theory at $e^{2}$-order
approximation. In order to avoid an awful reading, we place the most of
calculation in the Appendices, and some useful identities, as well. Our
remarks are given in Sec.6.


\section{Transition Amplitude}

To construct the transition amplitude, we must first do a
constraint analysis. Hence, before using the Faddeev-Senjanovic procedure,
we will present a short study, showing the main points of Hamiltonian
structure of $GQED_{4}$: the evaluation of canonical momenta,
followed by the determination of first- and second-class constraints, and, at last,
the choice of an appropriated set of gauge conditions. However, it will be
necessary to use the Faddeev-Popov-DeWitt method to get a convariant
expression to the transition amplitude. Thus, we start with the Lagrangian
density of $GQED_{4}$, defined by \footnote{%
We shall adopt, here, the metric convention $\eta _{\mu \nu }=diag.(+,-,-,-)$%
; the Greek and Latin indices runs from $0$ to $3$ and $1$ to $3$,
respectively, and the spinorial indices are represented by capital Latin
letters.}
\begin{equation}
\mathcal{L}=\frac{i}{2}\left( \bar{\psi}\hat{\partial}\psi -\bar{\psi}%
\overleftarrow{\hat{\partial}}\psi \right) -m\bar{\psi}\psi +e\bar{\psi}\hat{%
A}\psi -\frac{1}{4}F_{\mu \nu }F^{\mu \nu }+\frac{a^{2}}{2}\partial _{\mu
}F^{\alpha \mu }\partial ^{\beta }F_{\alpha \beta },  \label{I.1}
\end{equation}%
which, at classical level, is invariant under the local gauge transformations
\begin{equation}
\psi ^{\prime }\left( x\right) =e^{i\lambda \left( x\right) }\psi \left(
x\right) ,~\ A_{\mu }^{\prime }\left( x\right) =A_{\mu }\left( x\right) +%
\frac{1}{e}\partial _{\mu }\lambda \left( x\right) .  \label{I.2}
\end{equation}%
In the Lagrangian (\ref{I.1}), we used the following definitions: the
field-strengh tensor $F_{\nu \mu }\equiv \partial _{\nu }A_{\mu }-\partial
_{\mu }A_{\nu }$ and $\hat{O}\equiv \gamma _{\mu }O^{\mu }$. The Lagrangian $%
\mathcal{L}$ preserves all symmetries of usual $QED$. The Euler-Lagrange
equations following from the Hamiltonian principle with the corresponding
boundary conditions are
\begin{equation}
\left( i\hat{\partial}+e\hat{A}-m\right) \psi =0,~\left( 1+a^{2}\square
\right) \partial _{\mu }F^{\alpha \mu }=e\bar{\psi}\gamma ^{\alpha }\psi .
\label{I.3}
\end{equation}%
The canonical momenta, $\pi ^{\beta }$ and $\phi ^{\beta }$, conjugate to $%
A_{\alpha }$ and $\Gamma _{\alpha }$, respectively, where $\Gamma _{\alpha
}\equiv \partial _{0}A_{\alpha }$ are considered as independent variables,
defined \cite{13}, and given by
\begin{eqnarray}
\pi ^{\mu } &\equiv &\frac{\partial \mathcal{L}}{\partial \Gamma _{\mu }}%
-\partial _{0}\frac{\partial \mathcal{L}}{\partial \left( \partial
_{0}\Gamma _{\mu }\right) }-2\partial _{k}\frac{\partial \mathcal{L}}{%
\partial \left( \partial _{k}\Gamma _{\mu }\right) }=F^{\mu 0}-a^{2}\left[
\eta ^{\mu k}\partial _{k}\partial _{\lambda }F^{0\lambda }-\partial
_{0}\partial _{\lambda }F^{\mu \lambda }\right] ,  \label{I.4} \\
\phi ^{\mu } &\equiv &\frac{\partial \mathcal{L}}{\partial \left( \partial
_{0}\Gamma _{\mu }\right) }=a^{2}\left[ \smallskip \eta ^{\mu 0}\partial
_{\lambda }F^{0\lambda }-\partial _{\lambda }F^{\mu \lambda }\right] .
\label{I.4a}
\end{eqnarray}%
The canonical momenta associated with the fermion fields $\psi $ and $\bar{%
\psi}$ are
\begin{eqnarray}
p_{A} &\equiv &\frac{\partial \mathcal{L}}{\partial \left( \partial _{0}\bar{%
\psi}_{A}\right) }=\frac{i}{2}\left( \gamma ^{0}\psi \right) _{A},
\label{I.5a} \\
\bar{p}_{A} &\equiv &\frac{\partial \mathcal{L}}{\partial \left( \partial
_{0}\psi _{A}\right) }=\frac{i}{2}\left( \bar{\psi}\gamma ^{0}\right) _{A},
\label{I.5b}
\end{eqnarray}%
From the above momentum expressions, we shall study the constraint structure
of the theory following the Dirac's approach to singular systems \cite{14}.
From equations (\ref{I.4})-(\ref{I.5b}) and the linear independence of
constraints \cite{14}, it is possible to obtain the following set of first-class constraints:
\begin{equation}
\Omega _{1}\equiv \phi _{0}\approx 0,~\Omega _{2}\equiv \pi _{0}-\partial
_{k}\phi ^{k}\approx 0,~\Omega _{3}\equiv \partial _{k}\pi ^{k}+e\bar{\psi}%
\gamma ^{0}\psi \approx 0,  \label{I.6}
\end{equation}%
and a set of second-class ones,
\begin{equation}
\chi _{A}\equiv p_{A}-\frac{i}{2}\left( \gamma ^{0}\psi \right) _{A}\approx
0,~\bar{\chi}_{A}\equiv \bar{p}_{A}-\frac{i}{2}\left( \bar{\psi}\gamma
^{0}\right) _{A}\approx 0,  \label{I.7}
\end{equation}%
where "$\approx $" represents the fact that the relations (\ref{I.6}) and (%
\ref{I.7}) are weak equations, according to Dirac's procedure. The constraint
analysis presented here is justified by Faddeev-Senjanovic procedure to get the transition amplitude \cite{16}. This point will become clear
below.

The transition amplitude in the Hamiltonian form is written in the following way
\begin{equation}
Z=\int D\mu \exp \Big(i\int d^{4}x\Big[\pi ^{\mu }\left( \partial _{0}A_{\mu
}\right) +\phi _{\alpha }\left( \partial _{0}\Gamma ^{\alpha }\right)
-\left( \partial _{0}\psi \right) \bar{p}-\left( \partial _{0}\bar{\psi}%
\right) p-\mathcal{H_{C}}\Big]\Big),  \label{I.8}
\end{equation}%
where the canonical hamiltonian $\mathcal{H_{C}}$\ is given by
\begin{eqnarray}
\mathcal{H}_{\mathcal{C}} &=&\pi _{0}\Gamma ^{0}+\pi _{j}\Gamma ^{j}+\frac{%
\phi _{l}\phi ^{l}}{2a^{2}}+\phi _{l}\partial ^{l}\Gamma _{0}+\phi
_{l}\partial _{k}F^{lk}-\frac{i}{2}\bar{\psi}\gamma ^{j}\overleftrightarrow{%
\partial }_{j}\psi +m\bar{\psi}\psi -e\bar{\psi}\hat{A}\psi +\frac{1}{4}%
F_{kj}F^{kj}  \notag \\
&&+\frac{1}{2}\left( \Gamma _{j}-\partial _{j}A_{0}\right) ^{2}-\frac{a^{2}}{%
2}\left( \partial ^{j}\Gamma _{j}-\partial _{j}\partial ^{j}A_{0}\right)
^{2}.  \label{I.9a}
\end{eqnarray}%
The integration measure is defined by
\begin{equation}
D\mu =D\phi _{\nu }D\Gamma ^{\nu }D\pi ^{\mu }DA_{\mu }D\bar{\psi}D\psi D%
\bar{p}Dp\delta \left( \Theta _{l}\right) \det \left\vert \left\vert \left\{
\Omega _{a},\Sigma _{b}\right\} _{B}\right\vert \right\vert ~\det \left\vert
\left\vert \left\{ \chi _{A},\bar{\chi}_{B}\right\} _{B}\right\vert
\right\vert ^{-1/2}  \label{I.10}
\end{equation}%
where $\Theta =\{\Omega ,\Sigma ,\chi ,\bar{\chi}\}$, and the functionals $%
\Sigma $ are the gauge conditions that fix the first-class constraints.
Here, we will use the generalized radiation gauge condition
\begin{equation}
\Sigma _{1}\equiv \Gamma _{0}\approx 0,~\Sigma _{2}\equiv A_{0}\approx
0,~\Sigma _{3}\equiv \left( 1+a^{2}\square \right) \partial ^{k}A_{k}\approx
0,  \label{I.11}
\end{equation}%
which as it is shown in \cite{13}, is an appropriated set of noncovariant
gauge conditions for the first-class constraints (\ref{I.6}). We notice that
the determinant associated with the second-class constraints, $\det
\left\vert \left\vert \left\{ \chi _{A},\bar{\chi}_{B}\right\}
_{B}\right\vert \right\vert $, does not contain field variables, and so it
can be absorbed in a normalization constant; we also show that the
determinant between the first-class constraints (\ref{I.6}) and the gauge
fixing conditions (\ref{I.11}) has the form
\begin{equation}
\det \left\vert \left\vert \left\{ \Omega _{\alpha },\Sigma _{\beta
}\right\} _{B}\right\vert \right\vert =-\left( 1+a^{2}\nabla ^{2}\right)
\nabla ^{2}.  \label{I.12}
\end{equation}%
Therefore, through the following manipulations--combining the equations (\ref%
{I.10}) and (\ref{I.12}), substituting them into (\ref{I.8}), and also
carrying out momenta integrals and field variables-- we find the following
expression for the transition amplitude:
\begin{equation}
Z=\int DA_{\mu }D\bar{\psi}D\psi \det \left\vert \left\vert -\left(
1+a^{2}\nabla ^{2}\right) \nabla ^{2}\right\vert \right\vert \delta \left[
\left( 1+a^{2}\square \right) \partial ^{k}A_{k}\right] \exp \Big(i\int
d^{4}x\mathcal{L}\Big).  \label{I.13a}
\end{equation}%
Although equation (\ref{I.13a}) is correct, the noncovariant form is
not good for calculation purposes. However, we can use the ansatz of
Faddeev-Popov-DeWitt \cite{17} to achieve the desired covariant form for the
transition amplitude. Then, choosing the generalized Lorenz gauge condition
\cite{13}
\begin{equation}
\Omega \left[ A\right] =\left( 1+a^{2}\square \right) \partial ^{\mu }A_{\mu
}=0,  \label{I.14}
\end{equation}%
we finally obtain a expression for the covariant vacuum-vacuum transition
amplitude
\begin{eqnarray}
Z =\int DA_{\mu }D\bar{\psi}D\psi \det \left\vert \left\vert -\left(
1+a^{2}\square \right) \square \right\vert \right\vert \times \exp \Big(%
i\int d^{4}x\Big[\bar{\psi}\left( i\hat{\partial}-m+e\hat{A}\right) \psi -%
\frac{1}{4}F_{\mu \nu }F^{\mu \nu } &&  \notag \\
+\frac{a^{2}}{2}\partial ^{\mu }F_{\mu \beta }\partial _{\alpha }F^{\alpha
\beta }-\frac{1}{2\xi }\left( \left( 1+a^{2}\square \right) \partial ^{\mu
}A_{\mu }\right) ^{2}\Big]\Big).&&  \label{I.15}
\end{eqnarray}%
In this covariant gauge choice, we see that the Faddeev-Popov-DeWitt
determinant does not contain field variables (the ghosts decouple from the
gauge fields), and so, it can be absorbed into a normalization constant.


\section{Schwinger-Dyson-Fradkin Equations}

There are a lot of ways to extract the physical content of quantum field models,
but the most elegant one is from the Green's functions using
functional derivatives, which is a natural way to obtain such functions. The
method of functional derivatives, which has been largely used by Schwinger,
among others \cite{18,19}, uses a generating functional from which all
of Green's functions can be obtained by functional differentiation. These
equations are also known as Schwinger-Dyson-Fradkin equations (SDFE), and
the motivation to construct the SDFE's is the non-perturbative information
that is provided for the theory. However, if we regard these equations only as a
source of obtaining formal expansions in powers of the coupling constant, we
shall obtain nothing new in comparison with perturbation theory. The problem
of finding an effective method of solving those equations not based on
perturbation theory is, at present, still far from any sort of satisfactory
solution.

In the present section, we will derive these relations for the photon and electron
fields, and also for the vertex function. The first step is to define the
generating functional
\begin{equation}
\mathcal{Z}\left[ \eta ,\bar{\eta},J_{\mu }\right] =\int D\mu \left( \psi ,%
\bar{\psi},A_{\mu }\right) \exp \left[ i\mathcal{S}_{eff}\right] ,
\label{II.1}
\end{equation}%
with the effective action given by
\begin{eqnarray*}
\mathcal{S}_{eff} =\int d^{4}x\Big[\bar{\psi}\left( i\hat{\partial}-m+e\hat{A%
}\right) \psi -\frac{1}{4}F_{\mu \nu }F^{\mu \nu }+\frac{a^{2}}{2}\partial
^{\mu }F_{\mu \beta }\partial _{\alpha }F^{\alpha \beta }-\frac{1}{2\xi }%
\left( \left( 1+a^{2}\square \right) \partial ^{\mu }A_{\mu }\right) ^{2} &&
\\
+\bar{\psi}\eta +\bar{\eta}\psi +A^{\mu }J_{\mu }\Big],&&
\end{eqnarray*}%
where $\bar{\eta}$, $\eta $ and $J_{\mu }$ are the sources [auxiliary
mathematical device] for the fermion $\psi $, anti-fermion $\bar{\psi}$ and
the gauge $A_{\mu }$ fields, respectively. Let us stress that the components
of fermionic fields $\left( \bar{\psi},~\psi \right) $ and their sources $%
\left( \eta ,~\bar{\eta}\right) $ are elements of the Grassmann algebra, and
that $A_{\mu }$ and its source $J_{\mu }$, are c-numbers. From the
generating functional (\ref{II.1}), all the physical quantities of the theory
can be obtained. Whenever possible, we will discuss the meaning of
expressions of $GQED_{4}$ and also its points of equivalence or inequivalence with
the known results of $QED_{4}$.


\subsection{Schwinger-Dyson-Fradkin equation for Photon Propagator}


We will derive and discuss, here, the properties of the complete expression of the
gauge-field propagator in interaction with electrons. First, to obtain the
corresponding photon \emph{SDFE} we need to solve the following equation:
\begin{equation}
0=\left[ \left. \frac{\delta \mathcal{S}_{eff}}{\delta A_{\mu }\left(
x\right) }\right\vert _{\frac{\delta }{\delta i\eta },-\frac{\delta }{\delta
i\bar{\eta}},\frac{\delta }{\delta iJ_{\mu }}}+J^{\mu }\left( x\right) %
\right] \mathcal{Z}\left[ \eta ,\bar{\eta},J_{\mu }\right] ,  \label{II.3}
\end{equation}%
which, after evaluating the first term, can be written as
\begin{eqnarray}
-J^{\mu }\left( x\right) =\left[ \square \eta ^{\mu \nu }-\left[ 1-\frac{1}{%
\xi }\left( 1+a^{2}\square \right) \right] \partial ^{\mu }\partial ^{\nu }%
\right] \left( 1+a^{2}\square \right) \frac{\delta W}{\delta J^{\nu }\left(
x\right) }+ie\frac{\delta W}{\delta \eta \left( x\right) }\gamma ^{\mu }%
\frac{\delta W}{\delta \bar{\eta}\left( x\right) } &&  \notag \\
+ie\frac{\delta }{\delta \eta \left( x\right) }\left( \gamma ^{\mu }\frac{%
\delta W}{\delta \bar{\eta}\left( x\right) }\right) . &&  \label{II.4}
\end{eqnarray}%
The last equation represents the compact form of the non-perturbative
equivalent to the Podolsky field equation, subject to an external source $%
J_{\mu }$. The functional $W$ present in (\ref{II.4}) is the generating
functional for the connected Green's functions $W\left[ \eta ,\bar{\eta}%
,J_{\mu }\right] $, which is defined by $W\left[ \eta ,\bar{\eta},J_{\mu }%
\right] =-i\ln \mathcal{Z}\left[ \eta ,\bar{\eta},J_{\mu }\right] $. We also
introduce the generating functional for one-particle irreducible ($1PI$)
Green's functions $\Gamma \left[ \bar{\psi},\psi ,A_{\mu }\right] $ through
the Legendre transformation
\begin{equation}
\Gamma \left[ \bar{\psi},\psi ,A_{\mu }\right] =W\left[ \eta ,\bar{\eta}%
,J_{\mu }\right] -\int d^{4}x\left( \bar{\psi}\eta +\bar{\eta}\psi +A^{\mu
}J_{\mu }\right) .  \label{II.5}
\end{equation}%
From the above definitions, we obtain expressions for $\left( \bar{\psi}%
,\psi ,A_{\mu }\right) $ in terms of $\left( \eta ,\bar{\eta},J_{\mu
}\right) $, and vice versa, being given by
\begin{eqnarray}
&&A_{\mu }=\frac{1}{i}\frac{\delta W}{\delta J^{\mu }},~\psi =\frac{1}{i}%
\frac{\delta W}{\delta \bar{\eta}},~\bar{\psi}=-\frac{1}{i}\frac{\delta W}{%
\delta \eta }  \label{II.5b} \\
&&  \notag \\
&&J_{\mu }=-\frac{\delta \Gamma }{\delta A^{\mu }},~\eta =-\frac{\delta
\Gamma }{\delta \bar{\psi}},~\bar{\eta}=\frac{\delta \Gamma }{\delta \psi }.
\label{II.5a}
\end{eqnarray}%
Assuming the case that the fermionic sources are null, equation (\ref%
{II.4}) is written as
\begin{equation}
\frac{\delta \Gamma }{\delta A_{\mu }\left( x\right) }=\left[ T^{\mu \beta }+%
\frac{1}{\xi }\left( 1+a^{2}\square \right) L^{\mu \beta }\right] \left(
1+a^{2}\square \right) \square A_{\beta }\left( x\right) +ie\frac{\delta }{%
\delta \eta _{A}\left( x\right) }\left( \gamma ^{\mu }\frac{\delta W}{\delta
\bar{\eta}\left( x\right) }\right) _{A},  \label{II.6}
\end{equation}%
where we have used the following set of projectors
\begin{equation}
T^{\alpha \beta }+L^{\alpha \beta }=\eta ^{\alpha \beta },~L^{\alpha \beta }=%
\frac{\partial ^{\alpha }\partial ^{\beta }}{\square }.  \label{II.7}
\end{equation}%
From identifying
\begin{equation}
\mathscr{S}\left( x,y\right) \equiv i\left. \frac{\delta ^{2}W[\eta ,\bar{%
\eta},J_{\mu }]}{\delta \eta \left( y\right) \delta \bar{\eta}\left(
x\right) }\right\vert _{\psi =\bar{\psi}=0},  \label{II.8}
\end{equation}%
as the complete electron propagator in an external field $A_{\mu }$, which
satisfies the following functional relation
\begin{equation}
i\int d^{4}z\mathscr{S}_{BC}\left( x,z\right) \frac{\delta ^{2}\Gamma }{%
\delta \psi _{C}\left( y\right) \delta \bar{\psi}_{D}\left( z\right) }%
=\delta _{BD}\delta \left( x-y\right) ,  \label{II.9}
\end{equation}%
we can express (\ref{II.6}) as
\begin{eqnarray}
\frac{\delta \Gamma }{\delta A_{\mu }\left( x\right) }=\left[ T^{\mu \beta }+%
\frac{1}{\xi }\left( 1+a^{2}\square \right) L^{\mu \beta }\right] \left(
1+a^{2}\square \right) \square A_{\beta }\left( x\right) +eTr\left( \gamma ^{\mu }\mathscr{S}\left( x,x\right) \right) .
\label{II.10}
\end{eqnarray}%
Now, differentiating (\ref{II.10}) with respect to $A_{\nu }\left( y\right) $
and setting $J_{\mu }\left( x\right) =0$, yields
\begin{eqnarray}
\frac{\delta ^{2}\Gamma }{\delta A_{\nu }\left( y\right) \delta A_{\mu
}\left( x\right) }=\left[ T^{\mu \beta }+\frac{1}{\xi }\left( 1+a^{2}\square
\right) L^{\mu \beta }\right] \square \left( 1+a^{2}\square \right) \delta
\left( x-y\right) &&  \notag \\
-ieTr\left( \gamma ^{\mu }\frac{\delta }{\delta A_{\nu }\left( y\right) }%
\left( \frac{\delta ^{2}\Gamma }{\delta \psi \left( x\right) \delta \bar{\psi%
}\left( x\right) }\right) ^{-1}\right) . &&  \label{II.11}
\end{eqnarray}%
The second term on right-hand side of (\ref{II.11}) can be evaluated
immediately, giving a simple expression
\begin{equation}
\frac{\delta }{\delta A_{\nu }\left( y\right) }\left( \frac{\delta
^{2}\Gamma }{\delta \psi \left( x\right) \delta \bar{\psi}\left( x\right) }%
\right) ^{-1}=e\int d^{4}ud^{4}w\mathscr{S}\left( u,x\right) \Gamma ^{\nu
}\left( w,u;y\right) \mathscr{S}\left( x,w\right) .  \label{II.12}
\end{equation}%
where we take into account the definition (\ref{II.8}) and have introduced
the complete electron-photon vertex function
\begin{equation}
e\Gamma _{\mu }\left( x,y;z\right) \equiv \left. \frac{\delta ^{3}\Gamma }{%
\delta A^{\mu }\left( z\right) \delta \psi \left( y\right) \delta \bar{\psi}%
\left( x\right) }\right\vert _{A=\psi =\bar{\psi}=0}.  \label{II.13}
\end{equation}%
Similar to the fermionic case (\ref{II.9}), the second derivative of $\Gamma %
\left[ \bar{\psi},\psi ,A_{\mu }\right] $ with respect to $A_{\mu }\left(
x\right) $, generates the inverse of the photon propagator $\mathscr{D}_{\mu
\nu }\left( x-y\right) $. From this fact, and substituting (\ref{II.11}) into
(\ref{II.12}), then follows the \emph{SDFE} for the inverse of the complete photon
propagator
\begin{equation}
\mathscr{D}_{\mu \nu }^{-1}\left( x-y\right) =\Pi _{\mu \nu }\left(
x,y\right) +\left[ T_{\mu \nu }+\frac{1}{\xi }\left( 1+a^{2}\square \right)
L_{\mu \nu }\right] \left( 1+a^{2}\square \right) \square \delta \left(
x-y\right)  \label{II.14}
\end{equation}%
where the functional $\Pi _{\mu \nu }$ is known as photon self-energy
function, and is defined as
\begin{equation}
\Pi _{\mu \nu }\left( x,y\right) =-ie^{2}\int d^{4}ud^{4}wTr\Big[\mathscr{S}%
\left( u,x\right) \gamma _{\mu }\mathscr{S}\left( x,w\right) \Gamma _{\nu
}\left( w,u;x\right) \Big],  \label{II.15}
\end{equation}%
the $(-1)$ factor comes from the fermionic loop in the usual way. The $\Pi
_{\mu \nu }$ tensor describes the interaction of a photon with the
electron-positron field, and this interaction consists of creation and
annihilation of virtual pairs. Equation (\ref{II.15}) in momentum
representation assumes the form
\begin{equation}
\Pi _{\mu \nu }\left( k\right) =-\frac{ie^{2}}{(2\pi )^{4}}\int d^{4}pTr\Big[%
\mathscr{S}\left( p\right) \gamma _{\mu }\mathscr{S}\left( p-k\right) \Gamma
_{\nu }\left( p,k;p-k\right) \Big].  \label{II.15a}
\end{equation}

From the expression (\ref{II.14}), we can compute the gauge-field propagator
in a perturbative way, order-by-order, in the coupling constant $e$. An explicit
calculation, analysis, and discussion at the lowest order of radiative correction
of the Green's functions will be presented into Sec.5. Indeed, since the
bosonic $1PI$ function and complete photon propagator satisfy the identity $%
\Gamma_{\mu \sigma} (k) \mathscr{D}^{\sigma \nu }\left(k\right)=-i\eta
_{\mu}^{\nu}$, it is possible to find the following solution to the complete
photon propagator:
\begin{equation}
i\mathscr{D}_{\mu \nu }\left( k\right) =\frac{\eta _{\mu \nu} - \frac{k_{\mu
}k_{\nu }}{k^{2}}}{k^{2}\left[ \Pi \left( k\right) +\left(
1-a^{2}k^{2}\right) \right] }+\frac{\xi }{k^{2}\left(1-a^{2}k^{2}\right) ^{2}%
}\frac{k_{\mu }k_{\nu }}{k^{2}},  \label{II.18}
\end{equation}
where $\Pi$ is called scalar polarization, which is introduced due Lorentz
invariance of $\Pi _{\mu \nu}$, it has the structure
\begin{equation}
\Pi _{\mu \nu }\left( k\right) \equiv \left( -\eta _{\mu \nu}k^{2}+k_{\mu
}k_{\nu }\right) \Pi \left( k\right).  \label{II.17}
\end{equation}
It should be noted that the expression (\ref{II.18}) shows that the $\Pi(k)$
function is related with the transverse pole of photon propagator in
momentum representation. The diagramatic representation of the SDFE for the
photon propagator (\ref{II.18}) is shown in the Fig. 1.

The photon propagator at lowest order in perturbation theory, i.e., taking $%
\Pi \left( k\right) =0$ on (\ref{II.18}), can be conveniently written as
\begin{eqnarray}
iD_{\mu \nu }\left( k\right) =\left[ \eta _{\mu \nu }-\left( 1-\xi \right)
\frac{k_{\mu }k_{\nu }}{k^{2}}\right] \frac{1}{k^{2}}-\left[ \eta _{\mu \nu
}-\left( 1-\xi \right) \frac{k_{\mu }k_{\nu }}{k^{2}-\frac{1}{a^{2}}}\right]
\frac{1}{k^{2}-\frac{1}{a^{2}}} &&  \notag \\
+\left( 1-2\xi \right) \frac{k_{\mu }k_{\nu }}{k^{2}\left( k^{2}-\frac{1}{%
a^{2}}\right) }-\frac{k_{\mu }k_{\nu }}{\left( k^{2}-\frac{1}{a^{2}}\right)
^{2}}. &&  \label{II.19}
\end{eqnarray}%
As it can be seen in (\ref{II.19}), the beauty of this expression is the
appearance of the second term on right-hand side, which is originated from the Podolsky term
and the generalized Lorenz condition. Note that this term has massive poles, $%
m^{2}=a^{-2}$, which leads to a cancellation of IR divergences that are
present in the first term, the Maxwell's term. Furthermore, the separation of massless (usually $QED_{4}$) and massive modes in the propagator
expression (\ref{II.19}) in general gauge $\xi $ is owing to the linearity of
fields in the gauge terms of the Lagrangian (\ref{I.1}). By the relation between the
Podolsky parameter and the mass of photons, it is possible to set a bound
value for the photons mass, once we evaluate the parameter $a$ \cite{11,24}.

\begin{figure}[tbp]
\begin{center}
\scalebox{0.6}{\includegraphics{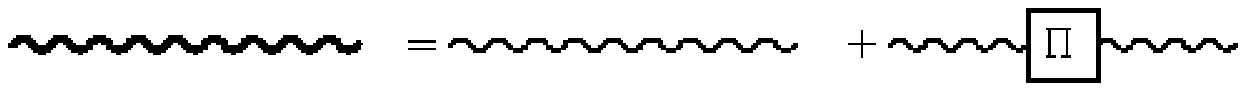}}
\end{center}
\par
\label{ESDFoton}
\caption{SDFE for the photon propagator}
\end{figure}

\subsection{Schwinger-Dyson-Fradkin equation for Fermionic Propagator}


In what follows in this subsection, we present the derivation of an integral
expression to the complete electron propagator $\mathscr{S}$. We also
introduce the mass operator $\mathscr{M}$, which contains all the radiative
correction to the motion of electron (in the same sense of polarization
operator $\Pi$ to the photons). We guide the derivation of SDFE for $%
\mathscr{S}$ in the same way as presented in last subsection for the photon
propagator. We recall that the functional equation
\begin{eqnarray}
0 &=&\left[ \left. \frac{\delta \mathcal{S}_{eff}}{\delta \bar{\psi}%
\left(x\right) }\right\vert _{\frac{\delta }{\delta i\eta },-\frac{\delta }{%
\delta i\bar{\eta}},\frac{\delta }{\delta iJ_{\mu }}}-\eta \left( x\right) %
\right] \mathcal{Z}\left[ \eta ,\bar{\eta},J_{\mu }\right]  \notag \\
&=&\eta \left( x\right) \mathcal{Z}+i\left(i\hat{\partial}-m-ie\gamma ^{\mu }%
\frac{\delta }{\delta J^{\mu }\left( x\right) }\right) \frac{\delta \mathcal{%
Z}}{\delta \bar{\eta}\left(x\right) } .  \label{II.20}
\end{eqnarray}
which is equivalent to the Dirac equation in presence of external sources,
will define a relation between $\mathscr{S}$ and $\mathscr{M}$. Now,
differentiating (\ref{II.20}) with respect to $\eta \left( y\right) $ and
taking the fermionic sources going to zero, one gets
\begin{equation}
i\delta \left( x-y\right) =\left( i\hat{\partial}-m+e\hat{A}(x)-ie\gamma
^{\mu }\frac{\delta }{\delta J^{\mu }\left( x\right) }\right) \mathscr{S}%
\left( x,y\right) .  \label{II.22}
\end{equation}
where we have used the definition (\ref{II.8}) for $\mathscr{S}$. Equation (\ref{II.22}) defines the non-perturbative connected two-point
fermionic Green's functions.

By means of a functional derivative identity, together with (\ref{II.5a})
and (\ref{II.12}), and also taking the source $J_{\mu }$ going to zero, the
last term of (\ref{II.22}) reads
\begin{equation}
\frac{\delta }{\delta J^{\mu }\left( x\right) }\mathscr{S}\left(
x,y;A\right) =e\int d^{4}ud^{4}zd^{4}w\mathscr{D}_{\mu \alpha }\left(
x-u\right) \mathscr{S}\left( x,w\right) \Gamma ^{\alpha }\left( w,z;u\right) %
\mathscr{S}\left( z,y\right) .  \label{II.23}
\end{equation}%
Also, note that the electromagnetic potential $A$ presented in the fourth
term of (\ref{II.22}) vanishes in the absence of an external source; that
is, $A_{\mu }\left( x;\ J_{\mu }=0\right) =0$. Combining this fact with (\ref%
{II.23}), the equation (\ref{II.22}) is rewritten as
\begin{equation}
i\delta \left( x-y\right) =\left( i\gamma ^{\mu }\partial _{\mu }-m\right) %
\mathscr{S}\left( x,y\right) -\int d^{4}z\Sigma (x,z)\mathscr{S}\left(
z,y\right)  \label{II.24}
\end{equation}%
where the electron self-energy operator $\Sigma $ introduced above is
defined by the following relation:
\begin{equation}
\Sigma (x-y)=ie^{2}\gamma ^{\mu }\int d^{4}ud^{4}w\mathscr{S}\left(
x,w\right) \mathscr{D}_{\mu \alpha }\left( u-x\right) \Gamma ^{\alpha
}\left( w,y;u\right) .  \label{II.24a}
\end{equation}%
which, on momentum representation, is written as
\begin{equation}
\Sigma (p)=\frac{ie^{2}}{(2\pi )^{4}}\gamma ^{\mu }\int d^{4}k\mathscr{S}%
\left( p-k\right) \mathscr{D}_{\mu \alpha }\left( k\right) \Gamma ^{\alpha
}\left( p,k;p-k\right) ,  \label{II.24b}
\end{equation}%
If we denote conveniently $\Sigma \mathscr{S}\left( x,y;A\right) =\int
d^{4}z\Sigma (x,z)\mathscr{S}\left( z,y;A\right) $, then we can rewrite
the equation (\ref{II.24}) in the following suitable form:
\begin{equation}
\left( i\gamma ^{\mu }\partial _{\mu }-m-\Sigma \right) \mathscr{S}\left(
x,y\right) =i\delta \left( x-y\right)  \label{II.25}
\end{equation}%
Moreover, introducing the so-called mass operator $\mathscr{M}$,
\begin{equation}
\mathscr{M}(x,y)=m\delta \left( x-y\right) +\Sigma (x,y)  \label{II.26}
\end{equation}%
into (\ref{II.25}), we obtain that the complete electron propagator in
momentum representation assumes the form
\begin{equation}
\mathscr{S}\left( p\right) =\frac{i}{\gamma ^{\mu }p_{\mu }-\mathscr{M}(p)}=%
\frac{i}{\gamma ^{\mu }p_{\mu }-m-\Sigma (p)}  \label{II.27}
\end{equation}%
which states the relation between the electron propagator and the mass
operator. The SDFE corresponding to the electron propagator is presented in
Fig. 2. Equations (\ref{II.25}) and (\ref{II.27}) show that the electron
propagator is the Green's function for an equation similar to the Dirac
equation $(\hat{p}-m-\Sigma )\psi =0$, but differing from the latter by the
addiction to the bare mass $m$ of the quantity $\Sigma $. For this reason, $\mathscr{M}$ is called \textit{mass operator}.

In a similar way to operator $\Pi$, we can say that the operator $\Sigma$
describes the interaction of the electron with its own electromagnetic
field. This interaction consists in emission and absorption of virtual
photons.

\begin{figure}[tbp]
\begin{center}
\scalebox{0.6}{\includegraphics{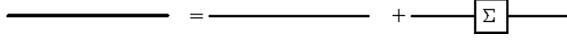}}
\end{center}
\caption{SDFE to electron propagator}
\end{figure}


\subsection{Schwinger-Dyson-Fradkin equation for Vertex}


As it is already known \cite{21}, it is impossible to construct for $QED_{4}$ a
closed integral equation that expresses the vertex function $\Gamma $ in
terms of $\mathscr{S}$ and $\mathscr{D}$ and that, together with the
equations (\ref{II.18}) and (\ref{II.27}), would give us a complete system of
equations determining the Green's functions. Nevertheless, it is
possible to find a relation connecting the vertex function $\Gamma $ with $%
\mathscr{S}$ and $\mathscr{D}$ \cite{20}; however, different from other Green's
functions, this relation contains only skeleton graphs \cite{19}, i.e.,
connected graphs. But, for our purposes here, it is enough to consider this
kind of approximation, due to the fact that, here, we have only interest in $%
e^{2}$-order calculation. Thus, recalling the vertex function is formally
obtained from
\begin{equation}
e\Gamma ^{\mu }(x,y;z)=\frac{\delta (\mathscr{S}(x,y;A))^{-1}}{\delta A(z)}
\label{II.28}
\end{equation}%
with $\mathscr{S}^{-1}$ being the inverse of the fermionic propagator (\ref{II.27}),
the vertex function can be also decomposed as
\begin{equation}
\Gamma _{\mu }(x,y;z)=-i\gamma _{\mu }\delta (x-y)\delta (y-z)+\Lambda _{\mu
}(x,y;z),  \label{II.29}
\end{equation}%
where $\Lambda _{\mu }$ is denoted as the vertex part of graphs. The vertex
function can be expressed in momentum space in terms of a new unknown
quantity, the electron-positron kernel $K$, by means of an integral equation
\cite{20}
\begin{eqnarray}
\Gamma _{\mu }(p,p^{\prime };k)=-i\gamma _{\mu }\delta (p+p^{\prime }-k)
&+&\int \frac{d^{4}q}{(2\pi )^{4}}[i\mathscr{S}(p^{\prime }+q)\Gamma _{\mu
}(q+p^{\prime },p+q)  \notag \\
&\times &i\mathscr{S}(p+q)]K(p+q,p^{\prime }+q,q),  \label{II.30}
\end{eqnarray}%
where $p^{\prime }$ and $p$ are, respectively, the momenta of the emerging and
incident electrons, while $k=p-p^{\prime }$ is the transferred momentum. $K$
consists of graphs with two external electron and two external positron
lines. Well, we have obtained, here, a closed integral equation for
the vertex function; however, for practical calculations we did not accomplished much, because $\Gamma _{\mu }$
is expressed in terms of an unknown quantity -- the kernel $K$. We shall
write down the complete kernel $K$ as a sum over skeleton graphs, which in
first-order yields \cite{20}
\begin{equation}
iK(p,p^{\prime },k)=(ie)^{2}\Gamma ^{\mu }(p,p-k)\mathscr{D}_{\mu \nu
}(k)\Gamma ^{\nu }(p^{\prime }-k,p^{\prime }).  \label{II.31}
\end{equation}%

Therefore, from (\ref{II.31}), we find out that the skeleton equation for the
vertex function (\ref{II.29}) written in Fourier representation is given by
\begin{eqnarray}
\Gamma _{\mu }(p,p^{\prime };k)=-i\gamma _{\mu }\delta (p-p^{\prime }-k) &+&%
\frac{e^{2}}{(2\pi )^{4}}\int d^{4}qi\mathscr{S}\left( p^{\prime }-q\right)
\Gamma _{\mu }\left( p-q,p^{\prime }-q;k\right) i\mathscr{S}\left( p-q\right)
\notag \\
&\times &\Gamma ^{\alpha }\left( p,p-q;q\right) i\mathscr{D}_{\nu \alpha
}\left( q\right) \Gamma ^{\nu }(p^{\prime },p^{\prime }-k,k).  \label{II.32}
\end{eqnarray}%
Figure 3 shows the vertex function. It is important to emphasize, here, that
the operators $\Sigma $, $\Pi $ and $\Lambda $ introduced above are
functional of the Green's functions $\mathscr{S}$, $\mathscr{D}$ and $\Gamma $,
which means that the self-energy functions are coupled, and one the Green's
function depends of another ones of lower order. Hence, we clearly see
that this tower of equations is related.

\begin{figure}[tbp]
\begin{center}
\scalebox{0.6}{\includegraphics{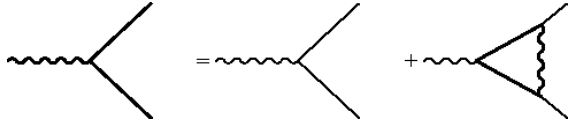}}
\end{center}
\caption{SDFE for the vertex function}
\end{figure}


\section{Ward-Fradkin-Takahashi Identities}


As it is well known, the generalized Ward-Fradkin-Takahashi (WFT) identities
are, in general, identities among Green's functions following from the
existence of a symmetry. The goal of this section is to derive these gauge
identities to $GQED_{4}$. First, we will show the WFT identity satisfied by $%
1PI$ gauge function, which leads to the transverse character of operator $%
\Pi _{\mu \nu }$. Next, we will derive the relation between the vertex
function and the inverse of the complete electron propagator, which is known as
the main WFT identity. At last, we will reproduce the main WFT identity in the $%
k\rightarrow 0$ limit (null transferred momentum). The derivation of these
identities is formally given as follows: starting from the generating
functional (\ref{II.1}) and performing the infinitesimal transformations
\begin{equation}
\psi ^{\prime }\left( x\right) =\psi \left( x\right) +i\lambda \left(
x\right) \psi \left( x\right) ,~A_{\mu }^{\prime }\left( x\right) =A_{\mu
}\left( x\right) +\frac{1}{e}\partial _{\mu }\lambda \left( x\right) ,
\label{IV.1}
\end{equation}%
and noticing that neither gauge fixing term nor the source terms are
invariant under these transformations, we find that the generating
functional $\mathcal{Z}\left[ \eta ,\bar{\eta},J^{\mu }\right] $ satisfie
the following equation of motion:
\begin{equation}
\Big[i\frac{\square }{e\xi }\left( 1+a^{2}\square \right) ^{2}\partial _{\mu
}\frac{\delta }{\delta J_{\mu }\left( x\right) }-\bar{\eta}\frac{\delta }{%
\delta \bar{\eta}\left( x\right) }+\eta \frac{\delta }{\delta \eta \left(
x\right) }-\frac{1}{e}\partial _{\mu }J^{\mu }\Big]\mathcal{Z}=0.
\label{IV.2}
\end{equation}%
The next step in deriving of WFT identities is to express (\ref{IV.2}) in
terms of the connected Green's functions $W\left[ \eta ,\bar{\eta},J^{\mu }%
\right] $ as
\begin{equation}
-\frac{\square }{e\xi }\left( 1+a^{2}\square \right) ^{2}\partial _{\mu }%
\frac{\delta W}{\delta J_{\mu }\left( x\right) }-i\bar{\eta}\frac{\delta W}{%
\delta \bar{\eta}\left( x\right) }+i\eta \frac{\delta W}{\delta \eta \left(
x\right) }-\frac{1}{e}\partial _{\mu }J^{\mu }=0.  \label{IV.3}
\end{equation}%
Finally, one can obtain the main quantum equation of motion for the theory by
writing (\ref{IV.3}) into an expression for the $1PI$ generating functional $%
\Gamma \left( \bar{\psi},\psi ,A_{\mu }\right) $ through (\ref{II.5}). Thus,
one has the general equation
\begin{equation}
-\frac{\square }{e\xi }\left( 1+a^{2}\square \right) ^{2}\partial _{\mu
}A^{\mu }(x)-i\frac{\delta \Gamma }{\delta \psi \left( x\right) }\psi \left(
x\right) +i\frac{\delta \Gamma }{\delta \bar{\psi}\left( x\right) }\bar{\psi}%
\left( x\right) +\frac{1}{e}\partial _{\mu }\frac{\delta \Gamma }{\delta
A_{\mu }\left( x\right) }=0.  \label{IV.4}
\end{equation}%
From equation (\ref{IV.4}), it is possible to derive all WFT identities.
Thus, the first identity comes by applying the functional derivative of $%
A_{\nu }\left( y\right) $ in equation (\ref{IV.4}) at $A_{\mu }=\psi =\bar{%
\psi}=0$
\begin{equation}
\partial _{\mu }\Gamma ^{\mu \nu }\left( x,y\right) -\frac{\square }{\xi }%
\left( 1+a^{2}\square \right) ^{2}\partial ^{\nu }\delta \left( x-y\right)
=0,  \label{IV.5}
\end{equation}%
which, together with equation (\ref{II.14}), imply in
\begin{equation}
k_{\mu }\Pi ^{\mu \nu }\left( k\right) =0.  \label{IV.6}
\end{equation}%
which shows the transverse character of operator $\Pi ^{\mu \nu }$. Now, the
main gauge WFT identity follows by taking the derivatives of expression (\ref%
{IV.4}) with respect to $\psi \left( y\right) $ and $\bar{\psi}(z)$ at $%
A_{\mu }=\psi =\bar{\psi}=0$, which in momentum space takes the form
\begin{equation}
k_{\mu }\tilde{\Gamma}^{\mu }\left( p,p^{\prime };k=p-p^{\prime }\right)
=i\left( 2\pi \right) ^{4}\left[ \mathscr{S}^{-1}\left( p\right) -\mathscr{S}%
^{-1}\left( p-p^{\prime }\right) \right] ,  \label{IV.8}
\end{equation}%
with $\mathscr{S}^{-1}$ the inverse of the complete electron propagator (%
\ref{II.27}).

Although the local gauge invariance at the classical level has been broken
in the quantum theory through the gauge fixing procedure and source terms,
the main WFT identity (\ref{IV.8}) holds, inheriting its essence, without
which the renormalizability cannot be guaranteed.

In the limit of null transferred momentum, i.e., $k\rightarrow 0$, the
equation (\ref{IV.8}) leads to a relation
\begin{equation}
i\tilde{\Gamma}^{\mu }\left( p,p;0\right) =\left( 2\pi \right) ^{4}\frac{%
\partial}{\partial p_{\mu}}\mathscr{S}^{-1}\left(p\right),  \label{IV.9}
\end{equation}
from this limit, also follows that
\begin{equation}
\tilde{\Lambda}^{\mu }\left( p,p;0\right) =-\frac{\partial}{\partial p_{\mu}}%
\Sigma \smallskip \left( p\right).  \label{IV.10}
\end{equation}
Both relations, (\ref{IV.9}) and (\ref{IV.10}), hold in the same way that they do
for $QED_{4}$.


\section{Radiative Corrections of the Second Order}


In the preceding sections, we have derived integral equations to the Green's
functions, the electron and photon propagators, and vertex function  for the $GQED_{4}$. Now, we will investigate the
corrections to these functions in the first nonvanishing order of
perturbation theory. The expression for the operator $\Pi_{\mu \nu}$ at $%
e^{2} $-order does not differ from that of the $QED_{4}$. This divergent result
implies, in the same way as for the $QED_{4}$ \cite{21}, the
renormalization of electronic charge $e$ and the introduction of
renormalization constant $Z_{3} $ in the $GQED_{4}$. Although the electron
self-energy function $\Sigma$ and vertex part $\Lambda$ in $e^{2}$-order are
different from the usual corrections for the $QED_{4}$, due to the presence of
Podolsky's terms in the free photon propagator $D_{\mu \nu}$ (\ref{II.19}), the
structure of divergences at this order, by power counting, remains the same
as $QED_{4}$, linearly and logarithmically divergent. At first glance, this
fact seems to lead to infinity results for the other two self-energy functions of
$GQED_{4}$ at $e^{2}$-order, thus, an explicit calculation of $\Sigma$ and $%
\Lambda$ expressions becomes necessary. These calculations are also
necessary to verify whether the main WFT identity (\ref{IV.8}) is still
satisfied at this order.

In order to use the dimensional regularization procedure, the Lagrangian
must have the right dimension (of internal loops); then, it is necessary
introduce the \textit{t'Hooft mass} $\mu$. Thus, also considering the case $%
\xi=1$, we have
\begin{equation}
\mathcal{L}_{eff}=\frac{i}{2}\bar{\psi}\overleftrightarrow{~\widehat{
\partial }}\psi -m\bar{\psi}\psi -e\mu ^{4-d} \bar{\psi}\widehat{A}\psi -%
\frac{1}{4} F_{\mu \nu }F^{\mu \nu }+ \frac{a^{2}}{2}\partial _{\mu }F^{\mu
\alpha}\partial _{\nu }F_{\alpha }^{\nu }-\frac{1}{2}\left[ \left(
1+a^{2}\square \right) \partial _{\mu }A^{\mu }\right] ^{2}.  \label{V.10}
\end{equation}
In the next two subsections, we shall compute the $\Sigma$ and $\Lambda$
functions. We will show that both functions, $\Sigma$ and $\Lambda$, can be
separated in two distinct contributions, the well-known contribution from the $%
QED_{4}$ and a new one that we will call Podolsky contribution. However,
we can observe by power counting that the Podolsky sector presents a
divergent share with the $QED$ sector; thus, we expect that they may cancel out
the divergence of the $GQED_{4}$. Now, we proceed to an explicit evaluation of the
electron self-energy and the vertex part.


\subsection{Electron Self-Energy}


We begin by investigating the second-order electron self-energy function.
This quantity corresponds to the diagram shown in Fig. 4.

In accordance with equation (\ref{II.24b}), the self-energy function $%
\Sigma $ can be written as
\begin{equation}
\Sigma ^{(2)}\left( p\right) \equiv \Sigma _{QED}\left( p\right) +\Sigma
_{Pod}\left( p\right)  \label{V.11}
\end{equation}%
where
\begin{equation*}
\Sigma _{QED}\left( p\right) =i\mu ^{4-d}\int \frac{d^{d}k}{\left( 2\pi
\right) ^{d}}\frac{1}{k^{2}}\gamma ^{\nu }\frac{\left( \widehat{p}-\widehat{k%
}+m\right) }{\left[ \left( p-k\right) ^{2}-m^{2}\right] }\gamma _{\nu }
\end{equation*}%
and
\begin{equation}
\Sigma _{Pod}\left( p\right) =-i\mu ^{4-d}\int \frac{d^{d}k}{\left( 2\pi
\right) ^{d}}\frac{\gamma ^{\lambda }\left( \widehat{p}-\widehat{k}+m\right)
}{\left[ \left( p-k\right) ^{2}-m^{2}\right] }\frac{\gamma ^{\nu }}{\left(
k^{2}-\frac{1}{a^{2}}\right) }\left[ \eta _{\lambda \nu }+\left[ \frac{1}{%
k^{2}}-\frac{1}{k^{2}-\frac{1}{a^{2}}}\right] k_{\lambda }k_{\nu }\right]
\label{V.12a}
\end{equation}

\begin{figure}[tbp]
\begin{center}
\scalebox{0.8}{\includegraphics{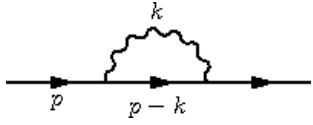}}
\end{center}
\caption{Electron self-energy diagram}
\end{figure}

The separation of $\Sigma ^{(2)}$ in two contributions is only made possible by
the linear structure of the free photon propagator (\ref{II.19}). First, the regularized $QED_{4}$ contribution for the electron
self-energy is given by \cite{21}:
\begin{equation}
\Sigma _{QED}\left( p\right) =\frac{1}{8\pi ^{2}}\frac{1}{\epsilon }\left(
\widehat{p}-4m\right) +\Sigma _{QED~Finite}\left( p\right)  \label{V.13}
\end{equation}%
with
\begin{equation}
\Sigma _{QED~Finite}\left( p\right) =-\frac{1}{16\pi ^{2}}\left[ \widehat{p}%
\left( \gamma +1\right) -2m\left( 2\gamma +1\right) \right] -\frac{1}{8\pi
^{2}}\int_{0}^{1}dz\left[ \widehat{p}\left( 1-z\right) -2m\right] \ln
\left\vert \frac{\left( 1-z\right) zp^{2}-m^{2}z}{4\pi \mu ^{2}}\right\vert
\label{V.13a}
\end{equation}%
where $\epsilon =4-d$ is the dimensional-regularization parameter.

Now, to evaluate the Podolsky contribution $\Sigma _{Pod}$ (\ref{V.12a}),
it is suitable to write it as
\begin{equation}
\Sigma _{Pod}\left( p\right) =\overset{3}{\underset{\alpha =1}{\sum }}\Sigma
_{Pod}^{\left( \alpha \right) }\left( p\right)  \label{V.14}
\end{equation}%
so that the quantities $\Sigma _{Pod}^{(i)}$ are defined by
\begin{equation}
\Sigma _{Pod}^{\left( 1\right) }\left( p\right) \equiv -i\mu ^{4-d}\int
\frac{d^{d}k}{\left( 2\pi \right) ^{d}}\frac{\gamma ^{\lambda }\left(
\widehat{p}-\widehat{k}+m\right) \gamma _{\lambda }}{\left[ \left(
p-k\right) ^{2}-m^{2}\right] }\frac{1}{\left( k^{2}-\frac{1}{a^{2}}\right) },
\label{V.15a}
\end{equation}%
\begin{equation}
\Sigma _{Pod}^{\left( 2\right) }\left( p\right) \equiv -i\mu ^{4-d}\int
\frac{d^{d}k}{\left( 2\pi \right) ^{d}}\frac{\widehat{k}\left( \widehat{p}-%
\widehat{k}+m\right) \widehat{k}}{\left[ \left( p-k\right) ^{2}-m^{2}\right]
}\frac{1}{k^{2}\left( k^{2}-\frac{1}{a^{2}}\right) },  \label{V.15b}
\end{equation}%
\begin{equation}
\Sigma _{Pod}^{\left( 3\right) }\left( p\right) \equiv i\mu ^{4-d}\int \frac{%
d^{d}k}{\left( 2\pi \right) ^{d}}\frac{\widehat{k}\left( \widehat{p}-%
\widehat{k}+m\right) \widehat{k}}{\left[ \left( p-k\right) ^{2}-m^{2}\right]
}\frac{1}{\left( k^{2}-\frac{1}{a^{2}}\right) ^{2}}.  \label{V.15c}
\end{equation}%
We are going now to calculate the expressions $\Sigma _{Pod}^{(i)}$, (\ref%
{V.15a})-(\ref{V.15c}). To solve conveniently the momentum
integration, we will use the Feynman parametrization and dimensional
regularization. Using the both procedures in equation (\ref{V.15a}), one
can put it in the form
\begin{equation*}
\Sigma _{Pod}^{\left( 1\right) }\left( p\right) =i\mu
^{4-d}\int_{0}^{1}dz\int \frac{d^{d}k}{\left( 2\pi \right) ^{d}}\frac{\gamma
^{\lambda }\left( \widehat{k}-\widehat{p}-m\right) \gamma _{\lambda }}{\left[
\left( k-pz\right) ^{2}+b^{2}\right] ^{2}}
\end{equation*}%
where $b^{2}=\left( 1-z\right) \left( zp^{2}-\frac{1}{a^{2}}\right) -m^{2}z$%
. Introducing the change of variables $k\rightarrow k-pz$, we obtain
\begin{equation}
\Sigma _{Pod}^{\left( 1\right) }\left( p\right) =-i\mu
^{4-d}\int_{0}^{1}dz\gamma ^{\lambda }\left[ \left( 1-z\right) \widehat{p}+m%
\right] \gamma _{\lambda }\int \frac{d^{d}k}{\left( 2\pi \right) ^{d}}\frac{1%
}{\left[ k^{2}+b^{2}\right] ^{2}}.  \label{V.17}
\end{equation}%
The k integration is carried out by using the identity (\ref{A.1c}), so that
(\ref{V.17}) reads
\begin{equation}
\Sigma _{Pod}^{\left( 1\right) }\left( p\right) =\left( -1\right) ^{\frac{d}{%
2}}\frac{\mu ^{4-d}}{\left( 4\pi \right) ^{\frac{d}{2}}}\Gamma \left( 2-%
\frac{d}{2}\right) \int_{0}^{1}dz\gamma ^{\lambda }\left[ \left( 1-z\right)
\widehat{p}+m\right] \gamma _{\lambda }[b^{2}]^{\frac{d}{2}-2}.  \label{V.18}
\end{equation}%
Now, expanding (\ref{V.18}) around $d=4$, we find that
\begin{equation}
\Sigma _{Pod}^{\left( 1\right) }\left( p\right) \underset{\epsilon
\rightarrow 0}{=}-\frac{1}{\epsilon }\frac{1}{8\pi ^{2}}\left( \widehat{p}%
-4m\right) +\Sigma _{Pod~Finite}^{\left( 1\right) }\left( p\right)
\label{V.19}
\end{equation}%
where
\begin{eqnarray}
\Sigma _{Pod~Finite}^{\left( 1\right) }\left( p\right) &=&\frac{1}{16\pi ^{2}%
}\left[ \widehat{p}\left( 1+\gamma \right) -2m\left( 1+2\gamma \right) %
\right]  \notag \\
&&+\frac{1}{8\pi ^{2}}\int_{0}^{1}dz\left[ \widehat{p}\left( 1-z\right) -2m%
\right] \ln \left\vert \frac{\left( 1-z\right) \left( zp^{2}-\frac{1}{a^{2}}%
\right) -m^{2}z}{4\pi \mu ^{2}}\right\vert .  \label{V.20}
\end{eqnarray}%
We can evaluate the other terms in a similar way; however, to avoid an
extensive calculus, we present here only the results, leaving the explicit
calculation of these quantities and other extensive expressions in \textit{%
Appendix B}. The evaluated expressions of them are
\begin{eqnarray}
\Sigma _{Pod}^{\left( 2\right) }\left( p\right) &=&\frac{1}{\epsilon }\frac{1%
}{8\pi ^{2}}\left( \smallskip m-\widehat{p}\right) +\Sigma
_{Pod~Finite}^{\left( 2\right) }\left( p\right)  \label{V.21a} \\
\Sigma _{Pod}^{\left( 3\right) }\left( p\right) &=&-\frac{1}{\epsilon }\frac{%
1}{8\pi ^{2}}\left( m-\widehat{p}\right) +\Sigma _{Pod~Finite}^{\left(
3\right) }(p)  \label{V.21}
\end{eqnarray}%
with the finite parts given by (\ref{B.5}) and (\ref{B.9}), respectively.

Indeed, by combining the results of equations (\ref{V.19}), (\ref{V.21a}) and (\ref{V.21})
into equation (\ref{V.14}), it follows that the regularized contribution of
the Podolsky sector for the electron self-energy function is given by
\begin{equation}
\Sigma _{Pod}\left( p\right) =-\frac{1}{8\pi ^{2}}\frac{1}{\epsilon }\left(%
\widehat{p}-4m\right) +\Sigma _{Pod~Finite}\left( p\right)  \label{V.22}
\end{equation}
where $\Sigma _{Pod~Finite}$ is given by (\ref{B.10}).

Therefore, it finally follows from a rearrangement of equations (\ref{V.13})
and (\ref{V.22}) that the electron self-energy function (\ref{V.11}), at $%
e^{2}$-order, has the following expression:
\begin{eqnarray}
\Sigma ^{(2)}\left( p\right) &=&\frac{1}{8\pi ^{2}}\int_{0}^{1}dz\left[
\left( 1-z\right) \widehat{p}-2m\right] \ln \left\vert \frac{\left(
1-z\right) zp^{2}-m^{2}z-\frac{1}{a^{2}}\left( 1-z\right) }{\left(
1-z\right) zp^{2}-m^{2}z}\right\vert  \notag \\
&+&\frac{1}{16\pi ^{2}}\int_{0}^{1}dx\int_{0}^{1-x}dy\left[ 2m-\left(
1+3y\right) \widehat{p}\right] \boldsymbol{A}_{1}\left( p,x,y\right)  \notag
\\
&+&\frac{1}{16\pi ^{2}}\int_{0}^{1}dx\int_{0}^{1-x}dy\left[ \left(
1-y\right) \widehat{p}+m\right] p^{2}y^{2}\boldsymbol{A}_{2}\left(
p,x,y\right)  \label{V.25}
\end{eqnarray}%
with the quantities $\boldsymbol{A}_{1}$ and $\boldsymbol{A}_{2}$ defined as
\begin{eqnarray}
\boldsymbol{A}_{1}\left( p,x,y\right) &\equiv &\ln \left\vert \frac{\left(
\smallskip 1-y\right) yp^{2}-m^{2}y-\frac{1}{a^{2}}\left( 1-y\right) }{%
\left( 1-y\right) yp^{2}-m^{2}y-\frac{x}{a^{2}}}\right\vert  \notag \\
\boldsymbol{A}_{2}\left( p,x,y\right) &\equiv &\frac{1}{\left( 1-y\right)
yp^{2}-m^{2}y-\frac{x}{a^{2}}}-\frac{1}{\left( 1-y\right) \left( yp^{2}-%
\frac{1}{a^{2}}\right) -m^{2}y}  \notag
\end{eqnarray}%
Equation (\ref{V.25}) shows that the electron self-energy function $%
\Sigma ^{(2)}$, at $e^{2}$-order, does not depends on $\mu $, and that it is
also free of divergences, which do not occurs in such ordinary $QED_{4}$ as
equation (\ref{V.13}). This last feature is an interesting property of the
theory. It seems that the Podolsky term in the Lagrangian (\ref{V.10}) acts
like a natural regulator of the theory, due to its massive character.
Nevertheless, a better analysis shows that the Podolsky term is not the
only one responsible for the finiteness of electron self-energy in $e^{2}$%
-order; the choice of the generalized Lorenz gauge condition (\ref{I.14}) is
also closely related to the finite result (\ref{V.25}). Hence, we can
conclude that the choice of the usual Lorenz condition to $GQED_{4}$ leads to
the divergent result for the self-energy of electron evaluated in the thesis advised
by Podolsky \cite{12}.


\subsection{Vertex Correction}


We now turn to the calculation of the vertex part $\Lambda^{\mu}(p^{\prime
},p;q=p-p^{\prime \mu}(p^{\prime },p)$ (\ref{II.32}), where, as usual, $%
p^{\prime }$ and $p$ are, respectively, the momenta of the emerging and
incident electron, while $q=p-p^{\prime }$ is the momentum of the incident
photon. The diagram that corresponds to this quantity is shown in Fig. 5.

In the same way that occurs in equation (\ref{V.11}) for the electron
self-energy function $\Sigma $, the vertex part (\ref{II.31}) also shows the
splitting of its expression in two distinct contributions:
\begin{equation}
\Lambda ^{\mu (2)}\left( p^{\prime },p\right) =\Lambda _{QED}^{\mu }\left(
p^{\prime },p\right) +\Lambda _{Pod}^{\mu }\left( p^{\prime },p\right).
\label{V.34}
\end{equation}%
One contribution comes from $QED_{4}$
\begin{equation}
\Lambda _{QED}^{\mu }\left( p^{\prime },p\right) =-i\mu ^{4-d}\int \frac{%
d^{d}k}{\left( 2\pi \right) ^{d}}\gamma ^{\alpha }\frac{\widehat{p}^{\prime
}-\widehat{k}+m}{\left( p^{\prime }-k\right) ^{2}-m^{2}}\gamma ^{\mu }\frac{%
\widehat{p}-\widehat{k}+m}{\left( p-k\right) ^{2}-m^{2}}\gamma _{\alpha }%
\frac{1}{k^{2}},  \label{V.35a}
\end{equation}%
and another one from Podolsky sector,
\begin{eqnarray}
\Lambda _{Pod}^{\mu }\left( p^{\prime },p\right) &=&i\mu ^{4-d}\int \frac{%
d^{d}k}{\left( 2\pi \right) ^{d}}\gamma ^{\alpha }\frac{\widehat{p}^{\prime
}-\widehat{k}+m}{\left( p^{\prime }-k\right) ^{2}-m^{2}}\gamma ^{\mu }\frac{%
\widehat{p}-\widehat{k}+m}{\left( p-k\right) ^{2}-m^{2}}\gamma ^{\beta }%
\frac{1}{k^{2}-\frac{1}{a^{2}}}  \notag \\
&&\times \left[ \eta _{\alpha \beta }+\left[ \frac{1}{k^{2}}-\frac{1}{k^{2}-%
\frac{1}{a^{2}}}\right] k_{\alpha }k_{\beta }\right] .  \label{V.35}
\end{eqnarray}%
The regularized $QED_{4}$ contribution (\ref{V.35a}) for the vertex
part is known as\cite{21}
\begin{equation}
\Lambda _{QED}^{\mu }\left( p^{\prime },p\right) =\frac{1}{\epsilon }\frac{1%
}{8\pi ^{2}}\gamma ^{\mu }+\Lambda _{QED~Finite}^{\mu }\left( p^{\prime
},p\right)  \label{V.36}
\end{equation}%
with $\Lambda _{QED~Finite}^{\mu }\left( p^{\prime },p\right) $ given by
\begin{equation}
\Lambda _{QED~Finite}^{\mu }\left( p^{\prime },p\right) =-\frac{1}{16\pi ^{2}%
}\int_{0}^{1}dx\int_{0}^{1-x}dy\frac{\mathbf{\Xi }^{\mu }\left( x,y,\widehat{%
p}^{\prime },\widehat{p}\right) }{\Delta ^{2}}-\frac{1}{8\pi ^{2}}\gamma
^{\mu }\Big[1+\frac{\gamma }{2}+\int_{0}^{1}dx\int_{0}^{1-x}dy\ln \left\vert
\frac{\Delta ^{2}}{4\pi \mu ^{2}}\right\vert \Big]  \label{V.36a}
\end{equation}%
where we have introduced the functions
\begin{eqnarray}
\mathbf{\Xi }^{\mu }\left( p^{\prime },p,x,y\right) &=&6\left( 1-x-y\right)
\widehat{p}^{\prime }\gamma ^{\mu }\widehat{p}+2mq_{\nu }[\gamma ^{\nu
},\gamma ^{\mu }]-4\left( 1-x-y+3xy\right) p.p^{\prime }\gamma ^{\mu
}+2m^{2}\gamma ^{\mu }  \notag \\
&+&2x\left( 1-x\right) \gamma ^{\mu }\widehat{p}^{2}+2y\left( 1-y\right)
\left( \widehat{p}^{\prime }\right) ^{2}\gamma ^{\mu }-4y\left( 1-y\right)
\widehat{p}^{\prime }\left( p^{\prime }\right) ^{\mu }-4x\left( 1-x\right)
\widehat{p}p^{\mu }  \notag \\
&-&4\left( 1-x-y-xy\right) (\widehat{p}^{\prime }p^{\mu }+\widehat{p}%
(p^{\prime })^{\mu })  \label{V.36b}
\end{eqnarray}%
and
\begin{equation}
\Delta ^{2}=xp^{2}\left( 1-x\right) +yp^{\prime 2}\left( 1-y\right)
-2xypp^{\prime }-m^{2}\left( x+y\right) ,  \label{V.37}
\end{equation}%
to simplify the notation of integrals.

Again, as it has happened with the Podolsky contribution for the electron
self-energy function $\Sigma _{Pod}$, the vertex part $\Lambda _{Pod}^{\mu }$
(\ref{V.35}) can also be written as three terms,
\begin{equation}
\Lambda _{Pod}^{\mu }\left( p^{\prime },p\right) =\overset{3}{\underset{%
\alpha =1}{\sum }}\Lambda _{Pod}^{\mu \left( \alpha \right) }\left(
p^{\prime },p\right)  \label{V.38}
\end{equation}%
where we have defined each term in the following way:
\begin{equation}
\Lambda _{Pod}^{\mu \left( 1\right) }\left( p^{\prime },p\right) \equiv i\mu
^{4-d}\int \frac{d^{d}k}{\left( 2\pi \right) ^{d}}\gamma ^{\alpha }\frac{%
\widehat{p}^{\prime }-\widehat{k}+m}{\left( p^{\prime }-k\right) ^{2}-m^{2}}%
\gamma ^{\mu }\frac{\widehat{p}-\widehat{k}+m}{\left( p-k\right) ^{2}-m^{2}}%
\gamma _{\alpha }\frac{1}{k^{2}-\frac{1}{a^{2}}},  \label{V.39a}
\end{equation}%
\begin{equation}
\Lambda _{Pod}^{\mu \left( 2\right) }\left( p^{\prime },p\right) \equiv i\mu
^{4-d}\int \frac{d^{d}k}{\left( 2\pi \right) ^{d}}\widehat{k}\frac{\widehat{p%
}^{\prime }-\widehat{k}+m}{\left( p^{\prime }-k\right) ^{2}-m^{2}}\gamma
^{\mu }\frac{\widehat{p}-\widehat{k}+m}{\left( p-k\right) ^{2}-m^{2}}%
\widehat{k}\frac{1}{k^{2}\left( k^{2}-\frac{1}{a^{2}}\right) },
\label{V.39b}
\end{equation}%
\begin{equation}
\Lambda _{Pod}^{\mu \left( 3\right) }\left( p^{\prime },p\right) \equiv
-i\mu ^{4-d}\int \frac{d^{d}k}{\left( 2\pi \right) ^{d}}\widehat{k}\frac{%
\widehat{p}^{\prime }-\widehat{k}+m}{\left( p^{\prime }-k\right) ^{2}-m^{2}}%
\gamma ^{\mu }\frac{\widehat{p}-\widehat{k}+m}{\left( p-k\right) ^{2}-m^{2}}%
\widehat{k}\frac{1}{\left( k^{2}-\frac{1}{a^{2}}\right) ^{2}}.  \label{V.39c}
\end{equation}

\begin{figure}[tbp]
\begin{center}
\scalebox{0.8}{\includegraphics{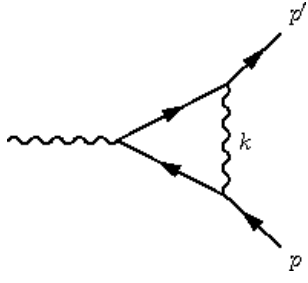}}
\end{center}
\caption{Vertex part diagram}
\end{figure}

To evaluate such integrals, we will proceed as we presented in the last
subsection for the electron self-energy function $\Sigma _{Pod}$. In this
subsection, we will only calculate one term, equation (\ref{V.39a}),
and present the results of another ones, (\ref{V.39b}) and (\ref{V.39c}),
leaving the calculation of the last two terms in \textit{Appendix C}.
Also, we exhibit there some extensive expressions that appear throughout
this subsection. Hence, recalling the Feynman parametrization, equation (\ref{V.39a}) can be expressed as
\begin{equation}
\Lambda _{Pod}^{\mu \left( 1\right) }\left( p^{\prime },p\right) =2i\mu
^{4-d}\int_{0}^{1}dx\int_{0}^{1-x}dy\int \frac{d^{d}k}{\left( 2\pi \right)
^{d}}\frac{\boldsymbol{N}^{\mu }\left( k,p^{\prime },p,x,y\right) }{\left[
k^{2}+\Delta ^{2}-\frac{1}{a^{2}}\left( 1-x-y\right) \right] ^{3}}
\label{V.40}
\end{equation}%
where we have replaced $k$ by $k-\smallskip xp-yp^{\prime }$ and defined the
function
\begin{equation}
\boldsymbol{N}^{\mu }\left( k,p^{\prime },p,x,y\right) \equiv \gamma
^{\sigma }\left[ \left( 1-y\right) \widehat{p}^{\prime }-\widehat{k}-x%
\widehat{p}+m\right] \gamma ^{\mu }\left[ \left( 1-x\right) \widehat{p}-%
\widehat{k}-y\widehat{p}^{\prime }+m\right] \gamma _{\sigma }  \notag
\end{equation}%
for convenience.

We now attend to the $k$ integration of (\ref{V.40}). Using the properties
of the Dirac matrices (\ref{A.6}) to separate the different $k$ terms in the
numerator and performing the momentum integration with the aid of
equations (\ref{A.1c}) and (\ref{A.1b}), we find that
\begin{eqnarray}
\Lambda _{Pod}^{\mu \left( 1\right) }\left( p^{\prime },p\right) &=&-\frac{%
\left( -1\right) ^{\frac{d}{2}}}{2}\frac{\mu ^{4-d}}{\left( 4\pi \right) ^{%
\frac{d}{2}}}\left( 2-d\right) ^{2}\gamma ^{\mu }\Gamma \left( 2-\frac{d}{2}%
\right) \int d\varsigma \frac{1}{\left[ \Delta ^{2}-\frac{1}{a^{2}}\left(
1-x-y\right) \right] ^{2-\frac{d}{2}}}  \notag \\
&&-\left( -1\right) ^{\frac{d}{2}}\frac{\mu ^{4-d}}{\left( 4\pi \right) ^{%
\frac{d}{2}}}\left( 2-d\right) \Gamma \left( 3-\frac{d}{2}\right) \int
d\varsigma \frac{\boldsymbol{M}^{\mu }\left( p^{\prime },p,x,y\right) }{%
\left[ \Delta ^{2}-\frac{1}{a^{2}}\left( 1-x-y\right) \right] ^{3-\frac{d}{2}%
}}  \notag \\
&&-\left( -1\right) ^{\frac{d}{2}}\frac{\mu ^{4-d}}{\left( 4\pi \right) ^{%
\frac{d}{2}}}\Gamma \left( 3-\frac{d}{2}\right) \int d\varsigma \frac{%
\mathbf{\Pi }^{\mu }\left( x,y,\widehat{p}^{\prime },\widehat{p}\right) }{%
\left[ \Delta ^{2}-\frac{1}{a^{2}}\left( 1-x-y\right) \right] ^{3-\frac{d}{2}%
}}  \label{V.41}
\end{eqnarray}%
with
\begin{equation}
\boldsymbol{M}^{\mu }\left( p^{\prime },p,x,y\right) =\left[ \left(
1-y\right) \widehat{p}^{\prime }-x\widehat{p}-m\right] \gamma ^{\mu }\left[
\left( 1-x\right) \widehat{p}-y\widehat{p}^{\prime }-m\right]
\end{equation}%
and
\begin{eqnarray}
\mathbf{\Pi }^{\mu }\left( x,y,\widehat{p}^{\prime },\widehat{p}\right) &=&4%
\left[ \left( 1-y\right) p^{\prime \sigma }-xp^{\sigma }\right] \left[
\left( 1-x\right) p_{\sigma }-yp_{\sigma }^{\prime }\right] \gamma ^{\mu }
\notag \\
&&-2\left[ \left( 1-y\right) \widehat{p}^{\prime }-x\widehat{p}-m\right] %
\left[ \left( 1-x\right) \widehat{p}-y\widehat{p}^{\prime }\right] \gamma
^{\mu }  \notag \\
&&-2\gamma ^{\mu }\left[ \left( 1-y\right) \widehat{p}^{\prime }-x\widehat{p}%
\right] \left[ \left( 1-x\right) \widehat{p}-y\widehat{p}^{\prime }-m\right]
\end{eqnarray}%
and the measure
\begin{equation}
\int d\varsigma \equiv \int_{0}^{1}dx\int_{0}^{1-x}dy.  \label{V.49b}
\end{equation}%
Equation (\ref{V.41}), expanded around $d=4$, gives the following expression:
\begin{equation}
\Lambda _{Pod}^{\mu \left( 1\right) ~}\left( p^{\prime },p\right) =-\frac{1}{%
\epsilon }\frac{1}{8\pi ^{2}}\gamma ^{\mu }+\Lambda _{Pod~Finite}^{\mu
\left( 1\right) ~}\left( p^{\prime },p\right)  \label{V.43}
\end{equation}%
with
\begin{equation}
\Lambda _{Pod~Finite}^{\mu \left( 1\right) }\left( p^{\prime },p\right) =%
\frac{1}{16\pi ^{2}}\int d\varsigma \frac{\mathbf{\Xi }^{\mu }\left(
p^{\prime },p,x,y\right) }{\left[ \Delta ^{2}-\frac{1}{a^{2}}\left(
1-x-y\right) \right] }+\frac{1}{8\pi ^{2}}\gamma ^{\mu }\Big[1+\frac{\gamma
}{2}+\int d\varsigma \ln \left\vert \frac{\Delta ^{2}-\frac{1}{a^{2}}\left(
1-x-y\right) }{4\pi \mu ^{2}}\right\vert \Big]  \label{V.44}
\end{equation}%
The evaluated expressions for $\Lambda _{Pod}^{\mu \left( 2\right) ~}\left(
p^{\prime },p\right) $ and $\Lambda _{Pod}^{\mu \left( 3\right) }\left(
p^{\prime },p\right) $ are [see \textit{Appendix C}]
\begin{eqnarray}
\Lambda _{Pod}^{\mu \left( 2\right) }\left( p^{\prime },p\right) &=&-\frac{1%
}{\epsilon }\frac{1}{8\pi ^{2}}\gamma ^{\mu }+\Lambda _{Pod~Finite}^{\mu
\left( 2\right) }\left( p^{\prime },p\right)  \label{V.45a} \\
\Lambda _{Pod}^{\mu \left( 3\right) }\left( p^{\prime },p\right) &=&\frac{1}{%
\epsilon }\frac{1}{8\pi ^{2}}\gamma ^{\mu }+\Lambda _{Pod~Finite}^{\mu
\left( 3\right) }\left( p^{\prime },p\right)  \label{V.45b}
\end{eqnarray}%
where the finite parts are given by (\ref{D.4}) and (\ref{D.7}),
respectively.

Therefore, when the equations (\ref{V.43}), (\ref{V.45a}) and (\ref%
{V.45b}) are combined, we determine the regularized expression for the Podolsky contribution
to the vertex part $\Lambda_{Pod}^{\mu }$ (\ref{V.38}):
\begin{eqnarray}
\Lambda _{Pod}^{\mu }\left( p^{\prime },p\right) &=&-\frac{1}{\epsilon }%
\frac{1}{8\pi ^{2}}\gamma ^{\mu }+\Lambda _{Pod~Finite}^{\mu }\left(
p^{\prime },p\right)  \label{V.46}
\end{eqnarray}
where $\Lambda _{Pod~Finite}^{\mu }$ is given by the equation (\ref{D.8}).

Substituting the results of the equations (\ref{V.36}) and (\ref{V.46}) into the definition (\ref%
{V.34}), we obtain that the vertex part $\Lambda ^{\mu }$ at $e^{2}$-order
has the following expression:
\begin{eqnarray}
\Lambda ^{\mu (2)}\left( p^{\prime },p\right) &=&\frac{1}{8\pi ^{2}}\gamma
^{\mu }\int d\varsigma \ln \left\vert \frac{\Delta ^{2}-\frac{1}{a^{2}}%
\left( 1-x-y\right) }{\Delta ^{2}}\right\vert +\frac{3}{8\pi ^{2}}\gamma
^{\mu }\int d\xi \ln \left\vert \frac{\Delta ^{2}-\frac{1}{a^{2}}z}{\Delta
^{2}-\frac{1}{a^{2}}\left( 1-x-y\right) }\right\vert  \notag \\
&&+\frac{1}{16\pi ^{2}}\int d\varsigma \mathbf{\Xi }^{\mu }\left( p^{\prime
},p,x,y\right) \left[ \frac{1}{\Delta ^{2}-\frac{1}{a^{2}}\left(
1-x-y\right) }-\frac{1}{\Delta ^{2}}\right]  \notag \\
&&+\frac{1}{16\pi ^{2}}\int d\xi \mathbf{\Sigma }_{\mathbf{1}}\left( x,y,%
\widehat{p}^{\prime },\widehat{p}\right) \gamma ^{\mu }\mathbf{\Sigma }_{%
\mathbf{2}}\left( x,y,\widehat{p}^{\prime },\widehat{p}\right) \left[ \frac{%
\mathbf{1}}{\left[ \Delta ^{2}-\frac{1}{a^{2}}\left( 1-x-y\right) \right]
^{2}}-\frac{\mathbf{1}}{\left[ \Delta ^{2}-\frac{1}{a^{2}}z\right] ^{2}}%
\right]  \notag \\
&&+\frac{1}{8\pi ^{2}}\int d\xi \Big[\mathbf{\Sigma }_{\mathbf{1}}\left( x,y,%
\widehat{p}^{\prime },\widehat{p}\right) \gamma ^{\mu }+\gamma ^{\mu }%
\mathbf{\Sigma }_{\mathbf{2}}\left( x,y,\widehat{p}^{\prime },\widehat{p}%
\right) -\frac{1}{4}\mathbf{\Sigma }_{\mathbf{3}}^{\mu }\left( x,y,\widehat{p%
}^{\prime },\widehat{p}\right)  \notag \\
&&+\frac{1}{2}\left( \widehat{p}^{\prime }-m\right) \gamma ^{\mu }\left(
\widehat{p}-m\right) \Big] \left[ \frac{\mathbf{1}}{\Delta ^{2}-\frac{1}{%
a^{2}}z}-\frac{\mathbf{1}}{\Delta ^{2}-\frac{1}{a^{2}}\left( 1-x-y\right) }%
\right]  \label{V.47}
\end{eqnarray}%
where we have defined the functions $\mathbf{\Sigma }_{\mathbf{1}}$,$\mathbf{%
\Sigma }_{\mathbf{2}}$ and $\mathbf{\Sigma }_{\mathbf{3}}^{\mu }$ as
\begin{equation*}
\mathbf{\Sigma }_{\mathbf{1}}\left( x,y,\widehat{p}^{\prime },\widehat{p}%
\right) =\left( x\widehat{p}+y\widehat{p}^{\prime }\right) \left[ \left(
1-y\right) \widehat{p}^{\prime }-x\widehat{p}+m\right] ,
\end{equation*}%
\begin{equation*}
\mathbf{\Sigma }_{\mathbf{2}}\left( x,y,\widehat{p}^{\prime },\widehat{p}%
\right) =\left[ \left( 1-x\right) \widehat{p}-y\widehat{p}^{\prime }+m\right]
\left( x\widehat{p}+y\widehat{p}^{\prime }\right) ,
\end{equation*}%
\begin{eqnarray}
\mathbf{\Sigma }_{\mathbf{3}}^{\mu }\left( x,y,\widehat{p}^{\prime },%
\widehat{p}\right) &=&-2\gamma ^{\mu }\left[ \left( 1-y\right) \widehat{p}%
^{\prime }-x\widehat{p}\right] \left( \widehat{p}-m\right) +4\left[ \left(
1-y\right) p^{\prime \sigma }-xp^{\sigma }\right] \left[ \left( 1-x\right)
p_{\sigma }-yp_{\sigma }^{\prime }\right] \gamma ^{\mu }  \notag \\
&&-2\left( \widehat{p}^{\prime }-m\right) \left[ \left( 1-x\right) \widehat{p%
}-y\widehat{p}^{\prime }\right] \gamma ^{\mu },  \label{V.48}
\end{eqnarray}%
and the measure
\begin{equation}
\int d\xi \equiv \int_{0}^{1}dx\int_{0}^{1-x}dy\int_{0}^{1-x-y}dz,
\label{V.49}
\end{equation}%
to simplify the notation of integrals, as we have been done with the
functions $\mathbf{\Xi }$ and $\Delta $.

As stated in the beginning of this section, we have shown that both
radiative corrections, the electron self-energy and vertex part, are finite
at $e^{2}$-order. Equation (\ref{V.47}) indicates the independence of
the vertex part with the t'Hooft mass $\mu$, and again, as it has happened
with the electron self-energy function, the finiteness of $\Lambda ^{\mu }$
is due to the Podolsky term plus the choice of the generalized Lorenz
gauge condition. Another important point is that the finiteness of vertex
part $\Lambda^{\mu }$ and electron self-energy function $\Sigma$ implies that
the main WFT identity (\ref{IV.8}) is still satisfied at $e^{2}$-order.


\section{Remarks and conclusions}


In this paper, the effects of the Podolsky term in the quantum theory of electron
and photon interactions were analyzed. After a constraint analysis, the
covariant transition amplitude was derived with the aid of the Faddeev-Popov-DeWitt ansatz in the generalized Lorenz gauge condition. The choice of this
gauge is of great importance to the obtained results. Then, we proceeded by deriving the SDFE's of theory by functional methods, and three Green's
functions have been determined: the photon $\mathscr{D}_{\mu \nu}$ and electron $%
\mathscr{S}$ propagators and the vertex function $\Gamma_{\mu}$, equations (\ref{II.18}),
(\ref{II.27}) and (\ref{II.32}), respectively. Through these functions, we
introduced the self-energy functions that contain the radiative corrections
in all order in pertubation theory: the polarization tensor $\Pi_{\mu\nu}$, the
mass operator $\mathscr{M}$, and the vertex part $\Lambda_{\mu}$,
respectively. However, modifications of these expressions compared to the
ones for $QED_{4}$ were observed only in mass operator and vertex part,
resulting from the contributions of the Podolsky electrodynamic. Although
such modifications are presented in all Green's functions, only the photon
propagator (\ref{II.19}) presents changes at tree level. Moreover, the most
interesting feature of this expression is the fact that we could separate
the usual contribution for the $QED_{4}$ in a general gauge $\xi$ from one that
arises from the Podolsky theory, and that the IR divergences presented in $%
QED_{4}$ terms are suppressed by the massive terms of Podolsky contribution.

The derivation of WFT identities was also presented. The first identity (\ref%
{IV.5}) showed that the transverse character of the polarization tensor $%
\Pi_{\mu\nu}$ is also preserved in the $GQED_{4}$ as in $QED_{4}$. Immediately,
we found the main WFT identity that relates the $1PI$ vertex function and
the complete electron propagator. The main WFT identity (\ref{IV.8}) is
responsible for holding the essence of the gauge symmetry in quantum level,
without which the renormalizability of theory cannot be guaranteed.

The last part of the article was devoted to the analysis of what the
Podolsky contribution brings to the quantum theory at $e^{2}$-order in
perturbation theory. At this order of approximation, we verified that the
photon self-energy function is divergent, showing that, if we claim to the
renormalization theory, the eletronic charge needs to be renormalized. Now,
for the other two corrections, interesting features appeared, and the free
photon propagator performs an important role in these analysis, giving
origin to a splitting of the correction functions in two distinct
contributions: one from the usual $QED_{4}$ and another from the Podolsky
theory. This splitting makes it possible to study each contribution independently.
Thus, since the $QED_{4}$ contribution is well-known in literature, our task
here was to calculate the Podolsky contribution to the electron
self-energy function and to the vertex part. And, the obtained expressions
for Podolsky contribution $\Sigma_{Pod}$ and $\Lambda_{Pod}^{\mu}$, equations (\ref%
{V.22}) and (\ref{V.46}), respectively, present the same divergent terms of the $%
QED_{4}$, equations (\ref{V.13}) and (\ref{V.36}), but with oppositive signs,
showing, then, that at $e^{2}$-order the $\Sigma$ and $\Lambda^{\mu}$
functions are finite. Although, here, we restrict ourselves to the case $\xi=1$
, these results can be generalized. It is possible to show that for $%
\xi\neq1$, the divergences associated with the electron self-energy function
and vertex part of the $QED_{4}$ are also canceled by the Podolsky
contribution. And, as an immediate consequence of the finiteness of $\Sigma$
and $\Lambda^{\mu}$, we verified that the main WFT identity (\ref{IV.8})
keeps being satisfied.

As a final comment, the Podolsky parameter $a$, which appears in all the
expressions evaluated here as a free parameter (as the inverse of photon
mass), can have its range of values limited through applications of
Podolsky theory. For example, we can evaluate now the physical quantity $%
\bar{u}(p^{\prime})\Lambda ^{\mu }u(p)$ that is related to the form
factors $F_{1}(q^{2})$ and $F_{2}(q^{2})$ of electric charge $e$, and to
the anomalous magnetic moment of the electron, respectively. We expect to
set a bound limit to the Podolsky parameter $a$ through the use of precise experimental data from
the electron magnetic moment, by calculating the
form factor $F_{2}(q^{2})$ for $GQED_{4}$. This study is now under
development. We can also express the quantum theory in a more formal and
constructive method, through dispersion relations \cite{22}, which can give
more transparent results and, also, a direct evaluation of electron anomalous
momentum. Another interesting issue is the study of the gauge properties of
the propagators for the $GQED_{4}$, constructing and analyzing the Landau-Khalatnikov-Fradkin transformation \cite{23} for the theory.
As mentioned before, a renormalization process for the photon propagator is
necessary, due to the divergence present in the self-energy; although the divergence
is the same as the $QED_{4}$, the renormalization constant and, also, the running
coupling constant may differ from the results for the $QED_{4}$ due to the poles
from the photon propagator expression (\ref{II.19}).

\noindent Going beyond of $T=0$, we can study the $GQED_{4}$ at finite
temperature, and derive all the thermodynamical quantities of theory,
including the energy-density distribution. And, following the idea of a
recent study of the Podolsky electromagnetism at finite temperature \cite%
{24}, where a bound value was set to the Podolsky parameter $a$ through the
energy distribution using the cosmic microwave background radiation
temperature, we can also use the cosmic microwave background radiation temperature to set a value to $a$
through the thermodynamical quantities of the $GQED_{4}$. These issues and
others will be further elaborated ,the subject of deep investigations, and
reported elsewhere.


\subsection*{Acknowledgements}


The authors would like to thank the referee for his/her comments and suggestions and
Professor A.T. Suzuki for carefully reading the manuscript and making suggestions.
R.B. thanks CNPq for full support, B.M.P. thanks CNPq for partial support and
G.E.R.Z. thanks VIPRIUDENAR for partial support.

\appendix

\section{$d$ Dimensions Identities}

As we made use of dimensional regularization in the evaluation of the radiative
correction expression, we present here some useful $d$-dimensional
identities associated with integrals, properties of gamma function, and Dirac
matrices.

\subsection{Integration in $d$-dimensions}


The useful results of integrals that appear throughout the paper are
\begin{equation}
\int \frac{d^{d}k}{\left( 2\pi \right) ^{d}}\frac{1}{\left(
k^{2}-m^{2}\right) ^{\alpha }}=\frac{i\left( -1\right) ^{\frac{d}{2}}}{%
\left( 4\pi \right) ^{\frac{d}{2}}}\frac{\Gamma \left( \alpha -\frac{d}{2}%
\right) }{\Gamma \left( \alpha \right) \left[ -m^{2}\right] ^{\alpha -\frac{d%
}{2}}}  \label{A.1c}
\end{equation}%
\begin{equation}
\int \frac{d^{d}k}{\left( 2\pi \right) ^{d}}\frac{k_{\mu }k_{\nu }}{\left(
k^{2}-m^{2}\right) ^{\alpha }}=\frac{i\left( -1\right) ^{\frac{d}{2}}\eta
_{\mu \nu }}{2\left( 4\pi \right) ^{\frac{d}{2}}}\frac{\Gamma \left( \alpha
-1-\frac{d}{2}\right) }{\Gamma \left( \alpha \right) \left[ -m^{2}\right]
^{\alpha -1-\frac{d}{2}}}  \label{A.1b}
\end{equation}%
\begin{equation}
\int \frac{d^{d}k}{\left( 2\pi \right) ^{d}}\frac{k_{\mu }k_{\nu }k_{\sigma
}k_{\rho }}{\left( k^{2}-m^{2}\right) ^{\alpha }}=\frac{i\left( -1\right) ^{%
\frac{d}{2}}}{4\left( 4\pi \right) ^{-\frac{d}{2}}}\Gamma \left( \alpha -2-%
\frac{d}{2}\right) \frac{\left[ \eta _{\mu \nu }\eta _{\sigma \rho }+\eta
_{\nu \rho }\eta _{\mu \sigma }+\eta _{\rho \mu }\eta _{\nu \sigma }\right]
}{\Gamma \left( \alpha \right) \left[ -m^{2}\right] ^{\alpha -2-\frac{d}{2}}}
\label{A.1a}
\end{equation}


\subsection{The Gamma Function}


An important property of the gamma function, with small $\epsilon $, is
given by the following relation:%
\begin{equation}
\Gamma \left( -n+\epsilon \right) =\frac{\left( -1\right) ^{n}}{n!}\left[%
\frac{1}{\epsilon }+\psi _{1}\left( n+1\right) +O\left( \epsilon \right) %
\right]  \label{A.2}
\end{equation}
where
\begin{equation}
\psi _{1}\left( n+1\right) =1+\frac{1}{2}+......+\frac{1}{n}-\gamma
\label{A.3}
\end{equation}%
and $\gamma $ is the Euler-Mascheroni constant. We needed the formulae
\begin{equation}
z\Gamma \left( z\right) =\Gamma \left( z+1\right),\quad \mathcal{X}^{-\frac{%
\epsilon }{2}}\simeq 1-\frac{\epsilon }{2}\ln\mathcal{X}  \label{A.4}
\end{equation}
as well.

\subsection{Dirac Matrices}


The algebra of Dirac matrices in $d$-dimensions is
\begin{equation}
\left\{ \gamma ^{\mu },\gamma ^{\nu }\right\} =2\eta ^{\mu \nu }  \label{A.5}
\end{equation}%
where $\eta ^{\mu \nu }$ is the metric tensor in $d$-dimensional Minkowski
space (with signature $+--...$), so that $\delta _{\mu }^{\mu }=d$; hence,
\begin{eqnarray}
\gamma ^{\mu }\gamma _{\mu } &=&d  \notag \\
\gamma ^{\sigma }\gamma ^{\mu }\gamma _{\sigma } &=&\left( 2-d\right) \gamma
^{\mu }  \notag \\
\gamma ^{\sigma }\gamma ^{\lambda }\gamma ^{\mu }\gamma _{\sigma }
&=&2\left( \gamma ^{\mu }\gamma ^{\lambda }-\gamma ^{\lambda }\gamma ^{\mu
}\right) +d\gamma ^{\lambda }\gamma ^{\mu }  \notag \\
\gamma ^{\alpha }\gamma ^{\sigma }\gamma ^{\mu }\gamma _{\sigma }\gamma
_{\alpha } &=&\left( 2-d\right) ^{2}\gamma ^{\mu }  \label{A.6} \\
\gamma ^{\alpha }\gamma ^{\sigma }\gamma ^{\mu }\gamma _{\alpha }\gamma
_{\sigma } &=&\left[ 2d-\left( \smallskip 2-d\right) ^{2}\right] \gamma
^{\mu }  \notag \\
\gamma ^{\sigma }\gamma ^{\lambda }\gamma ^{\mu }\gamma ^{\eta }\gamma
_{\sigma } &=&2\left( \gamma ^{\eta }\gamma ^{\lambda }\gamma ^{\mu }-\gamma
^{\mu }\gamma ^{\lambda }\gamma ^{\eta }+\gamma ^{\lambda }\gamma ^{\mu
}\gamma ^{\eta }\right) -d\gamma ^{\lambda }\gamma ^{\mu }\gamma ^{\eta }
\notag
\end{eqnarray}%
In addition
\begin{equation*}
Tr\left( \text{odd no. of }\gamma \text{ matrices}\right) =0
\end{equation*}%
\begin{equation*}
Tr~I=f\left( d\right) ,~Tr~\gamma _{\sigma }\gamma _{\alpha }=f\left(
d\right) \eta _{\sigma \alpha }
\end{equation*}%
\begin{equation*}
Tr~\gamma ^{\sigma }\gamma ^{\lambda }\gamma ^{\mu }\gamma ^{\eta }=f\left(
d\right) \left[ \eta ^{\sigma \lambda }\eta ^{\mu \eta }-\eta ^{\sigma \mu
}\eta ^{\lambda \eta }+\eta ^{\sigma \eta }\eta ^{\lambda \mu }\right]
\end{equation*}%
where $f\left( d\right) $ is an arbitrary well-behaved function, with $%
f\left( 4\right) =4$.


\section{Calculus of $\Sigma _{Pod}^{\left( 2\right) }\left( p\right) $ and $%
\Sigma _{Pod}^{\left( 3\right) }\left( p\right) $}


In order to evaluate the terms $\Sigma _{Pod}^{\left( 2\right) }$ and $%
\Sigma _{Pod}^{\left( 3\right) }$, we will follow the same steps presented
in the calculation of $\Sigma _{Pod}^{\left( 1\right) }$ in the subsection
5.1. First, we recall the Feynman parametrization and the dimensional
regularization. Thus, from (\ref{V.15b}), we obtain
\begin{equation}
\Sigma _{Pod}^{\left( 2\right) }\left( p\right) =-2i\mu ^{4-d}\int
d\varsigma \int \frac{d^{d}k}{\left( 2\pi \right) ^{d}}\frac{\left( \widehat{%
k}+\widehat{p}y\right) m\left[ \left( 1-y\right) \widehat{p}-\widehat{k}+m%
\right] \left( \widehat{k}+\widehat{p}y\right) }{\left[ k^{2}+b_{x}^{2}%
\right] ^{3}}  \label{B.1}
\end{equation}%
where we have changed $k\rightarrow k-py$, introduced $%
b_{x}^{2}=(1-y)yp^{2}-m^{2}+x\frac{1}{a^{2}}$, and used the equation (\ref%
{V.49b}) for the measure $d\varsigma $. Since the integral of the odd powers
of $k$ in numerator is zero, it is enough to evaluate the contribution of
the even powers. Then, carrying out the $k$ integration, the equation (\ref%
{B.1}) is written as
\begin{eqnarray}
\Sigma _{Pod}^{\left( 2\right) }\left( p\right) &=&\frac{\left( -1\right) ^{%
\frac{d}{2}}\mu ^{4-d}}{2\left( 4\pi \right) ^{\frac{d}{2}}}\Gamma \left( 2-%
\frac{d}{2}\right) \int d\varsigma \left[ \left[ 2\left( 1-y\right) -\left(
1+y\right) d\right] \widehat{p}+md\right] \left[ b_{x}^{2}\right] ^{\frac{d}{%
2}-2}  \notag \\
&&+\left( -1\right) ^{\frac{d}{2}}\frac{\mu ^{4-d}}{\left( 4\pi \right) ^{%
\frac{d}{2}}}\Gamma \left( 3-\frac{d}{2}\right) \int d\varsigma \left[
\left( 1-y\right) \widehat{p}+m\right] m^{2}y^{2}\left[ b_{x}^{2}\right] ^{%
\frac{d}{2}-3}  \label{B.3}
\end{eqnarray}%
Indeed, expanding (\ref{B.3}) for $d\rightarrow 4$, we find that $\Sigma
_{Pod}^{\left( 2\right) }$ can be expressed as
\begin{equation}
\Sigma _{Pod}^{\left( 2\right) }\left( p\right) =\frac{1}{\epsilon }\frac{1}{%
8\pi ^{2}}\left( m-\widehat{p}\right) +\Sigma _{Pod~Finite}^{\left( 2\right)
}\left( p\right)  \label{B.4}
\end{equation}%
where
\begin{eqnarray}
\Sigma _{Pod~Finite}^{\left( 2\right) }\left( p\right) &=&\frac{1}{16\pi ^{2}%
}\left[ \left( \gamma +\frac{2}{3}\right) \widehat{p}-\left( \gamma +\frac{1%
}{2}\right) m\right]  \notag \\
&&+\frac{1}{16\pi ^{2}}\int d\varsigma \Big [\frac{\left[ \left( 1-y\right)
\widehat{p}+m\right] p^{2}y^{2}}{b_{x}^{2}}-\left[ 2m-\left( 1+3y\right)
\widehat{p}\right] \ln \left\vert \frac{b_{x}^{2}}{4\pi \mu ^{2}}\right\vert %
\Big].  \label{B.5}
\end{eqnarray}%
The term $\Sigma _{Pod}^{\left( 3\right) }$ (\ref{V.15c}) is evaluated
following the same steps as in the previous ones through the Feynman parametrization and
dimensional regularization, and also replacing $k\rightarrow k-pz$:
\begin{equation}
\Sigma _{Pod}^{\left( 3\right) }\left( p\right) =2i\mu ^{4-d}\int d\varsigma
\int \frac{d^{d}k}{\left( 2\pi \right) ^{d}}\frac{\left( \widehat{k}+%
\widehat{p}y\right) \left[ \left( 1-y\right) \widehat{p}-\widehat{k}+m\right]
\left( \widehat{k}+\widehat{p}y\right) }{\left[ k^{2}+b^{2}\right] ^{3}},
\label{B.6}
\end{equation}%
where $b^{2}=\left( 1-y\right) \left( yp^{2}-\frac{1}{a^{2}}\right) -m^{2}y$%
, as we have defined in subsection 5.1. Carrying out the momentum
integration now, we find for (\ref{B.6}) the expression
\begin{eqnarray}
\Sigma _{Pod}^{\left( 3\right) }\left( p\right) &=&\left( -1\right) ^{\frac{d%
}{2}-1}\frac{\mu ^{4-d}}{2\left( 4\pi \right) ^{\frac{d}{2}}}\Gamma \left( 2-%
\frac{d}{2}\right) \int d\varsigma \left[ \left[ 2\left( 1-y\right) -\left(
1+y\right) d\right] \widehat{p}+md\right] \left[ b^{2}\right] ^{\frac{d}{2}%
-2}  \notag \\
&&-\left( -1\right) ^{\frac{d}{2}}\frac{\mu ^{4-d}}{\left( 4\pi \right) ^{%
\frac{d}{2}}}\Gamma \left( 3-\frac{d}{2}\right) \int d\varsigma \left[
\left( 1-y\right) \widehat{p}+m\right] p^{2}y^{2}\left[ b^{2}\right] ^{\frac{%
d}{2}-3},  \label{B.7}
\end{eqnarray}%
which in the limit $d\rightarrow 4$ is written as
\begin{equation}
\Sigma _{Pod}^{\left( 3\right) }\left( p\right) =-\frac{1}{\epsilon }\frac{1%
}{8\pi ^{2}}\left( m-\widehat{p}\right) +\Sigma _{Pod~Finite}^{\left(
3\right) }  \label{B.8}
\end{equation}%
with
\begin{eqnarray}
\Sigma _{Pod~Finite}^{\left( 3\right) }(p) &=&\frac{1}{16\pi ^{2}}\left[
\left( \gamma +\frac{1}{2}\right) m-\left( \gamma +\frac{2}{3}\right)
\widehat{p}\right]  \notag \\
&&-\frac{1}{16\pi ^{2}}\int d\varsigma \Big[\frac{\left[ \left( 1-y\right)
\widehat{p}+m\right] p^{2}y^{2}}{b^{2}}-\left[ 2m-\left( 1+3y\right)
\widehat{p}\right] \ln \left\vert \frac{b^{2}}{4\pi \mu ^{2}}\right\vert %
\Big]  \label{B.9}
\end{eqnarray}

Therefore, from the results os equations (\ref{V.20}), (\ref{B.5}) and (\ref{B.9}) we
obtain the following expression of the finite part of the Podolsky contribution $%
\Sigma _{Pod~Finite}$:
\begin{eqnarray}
\Sigma _{Pod~Finite} &=&\frac{1}{16\pi ^{2}}\left[ \left(\gamma +1\right)
\widehat{p}-2m\left( 2\gamma +1\right)\right] \notag \\
&&+ \frac{1}{16\pi ^{2}}\int d\varsigma \left[ \left[2m-\left( 1+3y\right) \widehat{p}\right] \ln \left|
\frac{(1-y)(yp^{2}-\frac{ 1}{a^{2}})-ym^{2}}{(1-y)yp^{2}-\frac{x}{a^{2}}%
-ym^{2} }\right| \right.  \notag \\
&&+\left. \widehat{p}\left[ \left( 1-y\right) \widehat{p}+m\right] \widehat{%
p }y^{2}\left[\frac{1}{(1-y)(yp^{2}-\frac{1}{a^{2}})-ym^{2}} -\frac{1}{
(1-y)yp^{2}-\frac{x}{a^{2}}-ym^{2}}\right] \right]  \notag \\
&&+\frac{1}{8\pi ^{2}}\int_{0}^{1}dz\left[ (1-z)\widehat{p}-2m\right] \ln
\left| \frac{(1-z)(zp^{2}-\frac{1}{a^{2}})-zm^{2}}{4 \pi \mu^{2}}\right|
\label{B.10}
\end{eqnarray}

\section{Calculus of $\Lambda _{Pod}^{\left( 2\right) ~\protect\mu %
}\left(p^{\prime },p\right) $ and $\Lambda _{Pod}^{\left( 3\right) ~\protect%
\mu }\left( p^{\prime },p\right) $}


In the same way as we did in Subsec. 5.2, we will proceed here into the
calculation of terms $\Lambda _{Pod}^{\mu \left( 2\right) }$ and $\Lambda
_{Pod}^{\mu \left( 3\right) }$ of the Podolsky contribution to the vertex part
at $e^{2}$-order. Now, recalling the Feynman parametrization and the
dimensional regularization, the equation (\ref{V.39b}) is expressed as
\begin{equation}
\Lambda _{Pod}^{\mu \left( 2\right) }\left( p^{\prime },p\right) =3i\mu
^{4-d}\left( 4\pi \right) ^{\frac{d}{2}}\int d\xi \int \frac{d^{d}k}{\left(
2\pi \right) ^{d}}\frac{\boldsymbol{O}^{\mu }\left( k,p^{\prime
},p,x,y\right) }{\left[ k^{2}+\Delta ^{2}-\frac{1}{a^{2}}z\right] ^{4}}
\label{D}
\end{equation}%
where we have replaced $k$ by $k-xp-yp^{\prime }$ and defined conveniently
the function
\begin{eqnarray}
\boldsymbol{O}^{\mu }\left( k,p^{\prime },p,x,y\right)  &=&(\widehat{k}-x%
\widehat{p}-y\widehat{p}^{\prime })\left[ \left( 1-y\right) \widehat{p}%
^{\prime }-\widehat{k}-x\widehat{p}+m\right] \gamma ^{\mu }\left[ \left(
1-x\right) \widehat{p}-\widehat{k}-y\widehat{p}^{\prime }+m\right]   \notag
\\
&&\times \gamma ^{\nu }(\widehat{k}-x\widehat{p}-y\widehat{p}^{\prime }).
\end{eqnarray}%
After a manipulation of matrices $\gamma $ in (\ref{D}) and evaluating the
momentum integration, one gets
\begin{eqnarray}
\Lambda _{Pod}^{\mu \left( 2\right) }\left( p^{\prime },p\right)  &=&-\frac{%
\left( -1\right) ^{\frac{d}{2}}}{4}\frac{\mu ^{4-d}}{\left( 4\pi \right) ^{%
\frac{d}{2}}}\left( d^{2}+2d\right) \gamma ^{\mu }\Gamma \left( 2-\frac{d}{2}%
\right) \int d\xi \frac{1}{\left[ \Delta ^{2}-\frac{1}{a^{2}}z\right] ^{2-%
\frac{d}{2}}}  \notag \\
&&-\left( -1\right) ^{\frac{d}{2}}\frac{\mu ^{4-d}}{\left( 4\pi \right) ^{%
\frac{d}{2}}}\Gamma \left( 4-\frac{d}{2}\right) \int d\xi \frac{\mathbf{\
\Sigma }_{\mathbf{1}}\left( x,y,\widehat{p}^{\prime },\widehat{p}\right)
\gamma ^{\mu }\mathbf{\Sigma }_{\mathbf{2}}\left( x,y,\widehat{p}^{\prime },%
\widehat{p}\right) }{\left[ \Delta ^{2}-\frac{1}{a^{2}}z\right] ^{4-\frac{d}{%
2}}}  \notag \\
&&+\frac{\left( -1\right) ^{\frac{d}{2}}}{2}\frac{\mu ^{4-d}}{\left( 4\pi
\right) ^{\frac{d}{2}}}\Gamma \left( 3-\frac{d}{2}\right) \int d\xi \frac{1}{%
\left[ \Delta ^{2}-\frac{1}{a^{2}}z\right] ^{3-\frac{d}{2}}}\Big[d\mathbf{%
\Sigma }_{\mathbf{1}}\left( x,y,\widehat{p}^{\prime },\widehat{p}\right)
\gamma ^{\mu }   \notag \\
&&+d\gamma ^{\mu }\mathbf{\Sigma }_{\mathbf{2}}\left( x,y,%
\widehat{p}^{\prime },\widehat{p}\right)-\mathbf{\Sigma }_{\mathbf{3}}^{\mu }\left( x,y,\widehat{p}^{\prime },%
\widehat{p}\right) -\left( 2-d\right) \left( \widehat{p}^{\prime }-m\right)
\gamma ^{\mu }\left( \widehat{p}-m\right) \Big]  \label{D.1}
\end{eqnarray}%
where we have defined the functions $\mathbf{\Sigma }_{\mathbf{1}}$, $%
\mathbf{\Sigma }_{\mathbf{2}}$ and $\mathbf{\Sigma }_{\mathbf{3}}^{\mu }$
and the measure $d\xi $ in (\ref{V.48}) and (\ref{V.49}), respectively.

Now, the equation (\ref{D.1}) in the limit $d\rightarrow 4$ assumes the form
\begin{equation}
\Lambda _{Pod}^{\mu \left( 2\right) }\left( p^{\prime },p\right) =-\frac{1}{%
\epsilon }\frac{1}{8\pi ^{2}}\gamma ^{\mu }+\Lambda _{Pod~Finite}^{\mu
\left( 2\right) }\left( p^{\prime },p\right)   \label{D.3}
\end{equation}%
with
\begin{eqnarray}
\Lambda _{Pod~Finite}^{\mu \left( 2\right) }\left( p^{\prime },p\right)  &=&%
\frac{1}{16\pi ^{2}}\gamma ^{\mu }\left[ \frac{5}{6}+\gamma +6\int d\xi \ln
\left\vert \frac{\Delta ^{2}-\frac{1}{a^{2}}z}{4\pi \mu ^{2}}\right\vert %
\right] \notag \\
&&-\frac{1}{16\pi ^{2}}\int d\xi \frac{\mathbf{\Sigma }_{\mathbf{1}%
}\left( x,y,\widehat{p}^{\prime },\widehat{p}\right) \gamma ^{\mu }\mathbf{%
\Sigma }_{\mathbf{2}}\left( x,y,\widehat{p}^{\prime },\widehat{p}\right) }{%
\left[ \Delta ^{2}-\frac{1}{a^{2}}z\right] ^{2}}  \notag \\
&&+\frac{1}{8\pi ^{2}}\int d\xi \frac{1}{\left[ \Delta ^{2}-\frac{1}{a^{2}}z%
\right] }\Big[\mathbf{\Sigma }_{\mathbf{1}}\left( x,y,\widehat{p}^{\prime },%
\widehat{p}\right) \gamma ^{\mu }+\gamma ^{\mu }\mathbf{\Sigma }_{\mathbf{2}%
}\left( x,y,\widehat{p}^{\prime },\widehat{p}\right)    \notag \\
&&-\frac{1}{4}\mathbf{\
\Sigma }_{\mathbf{3}}^{\mu }\left( x,y,\widehat{p}^{\prime },\widehat{p}%
\right)+\frac{1}{2}\left( \widehat{p}^{\prime }-m\right) \gamma ^{\mu }\left(
\widehat{p}-m\right) \Big]  \label{D.4}
\end{eqnarray}%
Following the same steps as before, we find for $\Lambda _{Pod}^{\mu \left(
3\right) }\left( p^{\prime },p\right) $, equation (\ref{V.39c}), the
following expression:
\begin{equation}
\Lambda _{Pod}^{\mu \left( 3\right) }\left( p^{\prime },p\right) =-3i\mu
^{4-d}\int d\xi \int \frac{d^{d}k}{\left( 2\pi \right) ^{d}}\frac{%
\boldsymbol{O}^{\mu }\left( k,p^{\prime },p,x,y\right) }{\left[ k^{2}+\Delta
^{2}-\frac{1}{a^{2}}(1-x-y)\right] ^{4}}  \label{D.a}
\end{equation}%
Now, when we evaluate the momentum integration in (\ref{D.a}), we get
\begin{eqnarray}
\Lambda _{Pod}^{\mu \left( 3\right) }\left( p^{\prime },p\right)  &=&\frac{%
\left( -1\right) ^{\frac{d}{2}}}{4}\frac{\mu ^{4-d}}{\left( 4\pi \right) ^{-%
\frac{d}{2}}}\left( d^{2}+2d\right) \gamma ^{\mu }\Gamma \left( 2-\frac{d}{2}%
\right) \int d\xi \frac{1}{\left[ \Delta ^{2}-\frac{1}{a^{2}}\left(
1-x-y\right) \right] ^{2-\frac{d}{2}}}  \notag \\
&&+\left( -1\right) ^{\frac{d}{2}}\frac{\mu ^{4-d}}{\left( 4\pi \right) ^{%
\frac{d}{2}}}\Gamma \left( 4-\frac{d}{2}\right) \int d\xi \frac{\mathbf{\
\Sigma }_{\mathbf{1}}\left( x,y,\widehat{p}^{\prime },\widehat{p}\right)
\gamma ^{\mu }\mathbf{\Sigma }_{\mathbf{2}}\left( x,y,\widehat{p}^{\prime },%
\widehat{p}\right) }{\left[ \Delta ^{2}-\frac{1}{a^{2}}\left( 1-x-y\right) %
\right] ^{4-\frac{d}{2}}}  \notag \\
&&-\frac{\left( -1\right) ^{\frac{d}{2}}}{2}\frac{\mu ^{4-d}}{\left( 4\pi
\right) ^{\frac{d}{2}}}\Gamma \left( 3-\frac{d}{2}\right) \int d\xi \frac{1}{%
\left[ \Delta ^{2}-\frac{1}{a^{2}}\left( 1-x-y\right) \right] ^{3-\frac{d}{2}%
}}\Big[d\mathbf{\Sigma }_{\mathbf{1}}\left( x,y,\widehat{p}^{\prime },%
\widehat{p}\right) \gamma ^{\mu }  \notag \\
&&+d\gamma ^{\mu }\mathbf{\Sigma }_{\mathbf{2}}\left( x,y,\widehat{p}%
^{\prime },\widehat{p}\right) -\mathbf{\Sigma }_{\mathbf{3}}^{\mu }\left(
x,y,\widehat{p}^{\prime },\widehat{p}\right) -\left( 2-d\right) \left(
\widehat{p}^{\prime }-m\right) \gamma ^{\mu }\left( \widehat{p}-m\right) %
\Big]  \label{D.5}
\end{eqnarray}%
which in the limit $d\rightarrow 4$ is expressed as
\begin{equation}
\Lambda _{Pod}^{\mu \left( 3\right) }\left( p^{\prime },p\right) =\frac{1}{%
\epsilon }\frac{1}{8\pi ^{2}}\gamma ^{\mu }+\Lambda _{Pod~Finite}^{\mu
\left( 3\right) }\left( p^{\prime },p\right)   \label{D.6}
\end{equation}%
with the finite part written as
\begin{eqnarray}
\Lambda _{Pod~Finite}^{\mu \left( 3\right) }\left( p^{\prime },p\right)  &=&-%
\frac{1}{16\pi ^{2}}\gamma ^{\mu }\left[ \frac{5}{6}+\gamma +6\int d\xi \ln
\left\vert \frac{\Delta ^{2}-\frac{1}{a^{2}}\left( 1-x-y\right) }{4\pi \mu
^{2}}\right\vert \right] \notag \\
&&+\frac{1}{16\pi ^{2}}\int d\xi \frac{\mathbf{\Sigma
}_{\mathbf{1}}\left( x,y,\widehat{p}^{\prime },\widehat{p}\right) \gamma
^{\mu }\mathbf{\Sigma }_{\mathbf{2}}\left( x,y,\widehat{p}^{\prime },%
\widehat{p}\right) }{\left[ \Delta ^{2}-\frac{1}{a^{2}}\left( 1-x-y\right) %
\right] ^{2}}  \notag \\
&&-\frac{1}{8\pi ^{2}}\int d\xi \frac{1}{\left[ \Delta ^{2}-\frac{1}{a^{2}}%
\left( 1-x-y\right) \right] }\Big[\mathbf{\Sigma }_{\mathbf{1}}\left( x,y,%
\widehat{p}^{\prime },\widehat{p}\right) \gamma ^{\mu }+\gamma ^{\mu }%
\mathbf{\Sigma }_{\mathbf{2}}\left( x,y,\widehat{p}^{\prime },\widehat{p}%
\right)   \notag \\
&&-\frac{1}{4}\mathbf{\Sigma }_{\mathbf{3}}^{\mu }\left( x,y,\widehat{p}%
^{\prime },\widehat{p}\right) +\frac{1}{2}\left( \widehat{p}^{\prime
}-m\right) \gamma ^{\mu }\left( \widehat{p}-m\right) \Big]  \label{D.7}
\end{eqnarray}%
After a rearrangement of expressions equations (\ref{V.44}), (\ref{D.4}) and (\ref{D.7}%
), we find that the expression of the finite part of Podolsky contribution $%
\Lambda _{Pod~Finite}^{\mu }$ is written as follows:
\begin{eqnarray}
\Lambda _{Pod~Finite}^{\mu }\left( p^{\prime },p\right)  &=&\frac{1}{8\pi
^{2}}\gamma ^{\mu }\left[ 1+\frac{\gamma }{2}+\int d\varsigma \ln \left\vert
\frac{\Delta ^{2}-\frac{1}{a^{2}}\left( 1-x-y\right) }{4\pi \mu ^{2}}%
\right\vert \right]   \notag \\
&&+\frac{1}{16\pi ^{2}}\int d\varsigma \frac{\mathbf{\Xi }^{\mu }\left(
p^{\prime },p,x,y\right) }{\Delta ^{2}-\frac{1}{a^{2}}\left( 1-x-y\right) }+%
\frac{3}{8\pi ^{2}}\gamma ^{\mu }\int d\xi \ln \left\vert \frac{\Delta ^{2}-%
\frac{1}{a^{2}}z}{\Delta ^{2}-\frac{1}{a^{2}}\left( 1-x-y\right) }%
\right\vert   \notag \\
&&+\frac{1}{16\pi ^{2}}\int d\xi \mathbf{\Sigma }_{\mathbf{1}}\left( x,y,%
\widehat{p}^{\prime },\widehat{p}\right) \gamma ^{\mu }\mathbf{\Sigma }_{%
\mathbf{2}}\left( x,y,\widehat{p}^{\prime },\widehat{p}\right) \left[ \frac{%
\mathbf{1}}{\left[ \Delta ^{2}-\frac{1}{a^{2}}\left( 1-x-y\right) \right]
^{2}}-\frac{\mathbf{1}}{\left[ \Delta ^{2}-\frac{1}{a^{2}}z\right] ^{2}}%
\right]   \notag \\
&&+\frac{1}{8\pi ^{2}}\int d\xi \Big[\mathbf{\Sigma }_{\mathbf{1}}\left( x,y,%
\widehat{p}^{\prime },\widehat{p}\right) \gamma ^{\mu }+\gamma ^{\mu }%
\mathbf{\Sigma }_{\mathbf{2}}\left( x,y,\widehat{p}^{\prime },\widehat{p}%
\right) -\frac{1}{4}\mathbf{\Sigma }_{\mathbf{3}}^{\mu }\left( x,y,\widehat{p%
}^{\prime },\widehat{p}\right)   \notag \\
&&+\frac{1}{2}\left( \widehat{p}^{\prime }-m\right) \gamma ^{\mu }\left(
\widehat{p}-m\right) \Big]\left[ \frac{1}{\Delta ^{2}-\frac{1}{a^{2}}z}-%
\frac{\mathbf{1}}{\Delta ^{2}-\frac{1}{a^{2}}\left( 1-x-y\right) }\right] .
\label{D.8}
\end{eqnarray}

\end{document}